\documentclass[twocolumn,showpacs,superscriptaddress,nofootinbib,floatfix]{revtex4}
\usepackage{graphicx}

\begin{document}
\def\Journal#1#2#3#4{{#1} {\bf #2}, #3 (#4)}
\title{\bf Charge Independent(CI) and Charge Dependent(CD) correlations as a
function of Centrality formed from $\Delta \phi \Delta \eta$ Charged Pair 
Correlations in Minimum Bias Au+Au Collisions at $\sqrt{s_{NN}} =$  200 GeV}
\affiliation{Argonne National Laboratory, Argonne, Illinois 60439}
\affiliation{University of Birmingham, Birmingham, United Kingdom}
\affiliation{Brookhaven National Laboratory, Upton, New York 11973}
\affiliation{California Institute of Technology, Pasadena, California 91125}
\affiliation{University of California, Berkeley, California 94720}
\affiliation{University of California, Davis, California 95616}
\affiliation{University of California, Los Angeles, California 90095}
\affiliation{Universidade Estadual de Campinas, Sao Paulo, Brazil}
\affiliation{Carnegie Mellon University, Pittsburgh, Pennsylvania 15213}
\affiliation{University of Illinois at Chicago, Chicago, Illinois 60607}
\affiliation{Creighton University, Omaha, Nebraska 68178}
\affiliation{Nuclear Physics Institute AS CR, 250 68 \v{R}e\v{z}/Prague, Czech Republic}
\affiliation{Laboratory for High Energy (JINR), Dubna, Russia}
\affiliation{Particle Physics Laboratory (JINR), Dubna, Russia}
\affiliation{University of Frankfurt, Frankfurt, Germany}
\affiliation{Institute of Physics, Bhubaneswar 751005, India}
\affiliation{Indian Institute of Technology, Mumbai, India}
\affiliation{Indiana University, Bloomington, Indiana 47408}
\affiliation{Institut de Recherches Subatomiques, Strasbourg, France}
\affiliation{University of Jammu, Jammu 180001, India}
\affiliation{Kent State University, Kent, Ohio 44242}
\affiliation{University of Kentucky, Lexington, Kentucky, 40506-0055}
\affiliation{Institute of Modern Physics, Lanzhou, China}
\affiliation{Lawrence Berkeley National Laboratory, Berkeley, California 94720}
\affiliation{Massachusetts Institute of Technology, Cambridge, MA 02139-4307}
\affiliation{Max-Planck-Institut f\"ur Physik, Munich, Germany}
\affiliation{Michigan State University, East Lansing, Michigan 48824}
\affiliation{Moscow Engineering Physics Institute, Moscow Russia}
\affiliation{City College of New York, New York City, New York 10031}
\affiliation{NIKHEF and Utrecht University, Amsterdam, The Netherlands}
\affiliation{Ohio State University, Columbus, Ohio 43210}
\affiliation{Panjab University, Chandigarh 160014, India}
\affiliation{Pennsylvania State University, University Park, Pennsylvania 16802}
\affiliation{Institute of High Energy Physics, Protvino, Russia}
\affiliation{Purdue University, West Lafayette, Indiana 47907}
\affiliation{Pusan National University, Pusan, Republic of Korea}
\affiliation{University of Rajasthan, Jaipur 302004, India}
\affiliation{Rice University, Houston, Texas 77251}
\affiliation{Universidade de Sao Paulo, Sao Paulo, Brazil}
\affiliation{University of Science \& Technology of China, Hefei 230026, China}
\affiliation{Shanghai Institute of Applied Physics, Shanghai 201800, China}
\affiliation{SUBATECH, Nantes, France}
\affiliation{Texas A\&M University, College Station, Texas 77843}
\affiliation{University of Texas, Austin, Texas 78712}
\affiliation{Tsinghua University, Beijing 100084, China}
\affiliation{Valparaiso University, Valparaiso, Indiana 46383}
\affiliation{Variable Energy Cyclotron Centre, Kolkata 700064, India}
\affiliation{Warsaw University of Technology, Warsaw, Poland}
\affiliation{University of Washington, Seattle, Washington 98195}
\affiliation{Wayne State University, Detroit, Michigan 48201}
\affiliation{Institute of Particle Physics, CCNU (HZNU), Wuhan 430079, China}
\affiliation{Yale University, New Haven, Connecticut 06520}
\affiliation{University of Zagreb, Zagreb, HR-10002, Croatia}

\author{B.I.~Abelev}\affiliation{University of Illinois at Chicago, Chicago, Illinois 60607}
\author{M.M.~Aggarwal}\affiliation{Panjab University, Chandigarh 160014, India}
\author{Z.~Ahammed}\affiliation{Variable Energy Cyclotron Centre, Kolkata 700064, India}
\author{B.D.~Anderson}\affiliation{Kent State University, Kent, Ohio 44242}
\author{D.~Arkhipkin}\affiliation{Particle Physics Laboratory (JINR), Dubna, Russia}
\author{G.S.~Averichev}\affiliation{Laboratory for High Energy (JINR), Dubna, Russia}
\author{Y.~Bai}\affiliation{NIKHEF and Utrecht University, Amsterdam, The Netherlands}
\author{J.~Balewski}\affiliation{Massachusetts Institute of Technology, Cambridge, MA 02139-4307}
\author{O.~Barannikova}\affiliation{University of Illinois at Chicago, Chicago, Illinois 60607}
\author{L.S.~Barnby}\affiliation{University of Birmingham, Birmingham, United Kingdom}
% ...

\collaboration{STAR Collaboration}\noaffiliation

\date{\today}

\begin{abstract}
We report high precision charged-particle pair (2-D) correlation analyses in
the space of $\Delta \phi$ (azimuth) and $\Delta \eta$ (pseudorapidity), for
minimum bias Au + Au collisions at $\sqrt{s_{NN}}$ = 200 GeV as a function of
centrality (0-80\%). The intermediate transverse momenta region chosen
$0.8 < p_t < 4.0$ GeV/c corresponds to an emission source size $\sim2$fm 
obtained from HBT measurements and should resolve substructures at the scale of
$\sim2$fm. The difference and the sum of unlike-sign and like-sign charged 
pairs form Charge Dependent (CD) correlations and Charge Independent (CI) 
correlations respectively. The CD displays the initial correlation at 
hadronization of the opposite sign pairs emitted from the same
space-time region as modified by further medium interactions before kinetic
freeze-out. Our analysis of the CD correlations shows approximately jet-like
structure, independent of centrality and is consistent with the initial 
correlation which is predicted by Pythia (or HIJING) jets. The CI correlation
displays the average structure of the correlated emitting sources after 
kinetic freeze-out. For the most central bins, the $\Delta \eta$ width of the 
CI correlation on the near side ($\Delta \phi$ around $0^\circ$) is elongated 
by a factor $\sim$3 destroying the jet-like symmetry. This elongation decreases
continually with decreasing centrality and essentially restores the jet-like
symmetric structure in the most peripheral bins. The Pythia and HIJING event
generators together with a QCD inspired Parton Bubble Model (PBM), which 
motivated this analysis, are used to compare to our data. We discuss the 
arguments for substructure, surface emission, and opacity in the central 
fireball region.
\end{abstract}
                                                                      

\maketitle
\section{introduction}
The Search for a Quark-Gluon Plasma (QGP) \cite{whitepaper, QuarkM}
has been a high priority task at the Relativistic Heavy Ion Collider, RHIC
\cite{RHIC}. Central Au + Au collisions at RHIC exceed \cite{transenergy}
the initial energy density that is predicted by lattice Quantum Chromodynamics
(QCD) to be sufficient for production of QGP \cite{LQCD}. Observations of
substructure have historically played an important role in advancing
scientific progress in nuclear and particle physics. Correlations are a
powerful tool in the search for substructures. The correlations generated by
particle pairs (e.g. \cite{HBT,mikhail,aya,centralproduction}) have been 
investigated. For central Au + Au collisions one might expect particle pair 
correlations to be reduced by thermalization. However, correlations 
could come from two possible sources. One source could be detectable 
correlations from substructures which form on the surface of the 
fireball at kinetic freeze-out \cite{PBM,themodel,VanHove,LL}. The 
other is correlations from initial hard scatterings that have only been 
modified by interactions with the medium of the fireball
interior but not destroyed \cite{aya}\footnote{There are different types of 
models employed}. Various models are discussed and comparisons with analyses
are made in Section VI.
                                                  
In this paper we analyze the Charge Independent (CI) and Charge Dependent (CD)
correlations in the two dimensional space --- $\Delta \phi$ and $\Delta \eta$
--- of charged pairs resulting from minimum bias Au + Au collisions at
$\sqrt{s_{NN}}$ = 200 GeV. The correlations were formed by particles in the
intermediate $p_t$ range 0.8 $< p_t <$ 4.0 GeV/c. The two types of charge 
pairs are the unlike-sign (US) and the like-sign (LS). The total 
correlations which are physically significant are CI = US + LS and 
CD = US - LS (Section IV and VI). If the background (equation 2 of Section 
III C) is subtracted from these correlations we obtain the corresponding 
signal correlations (CI signal and CD signal).
                                                                      
We have performed the present analyses using a model independent method so 
that these intermediate $p_t$ range correlation results can be theoretically 
analyzed in any way. However, in order to extract the signals we are 
interested in, we assume that the background is composed of known, 
expected, and instrumental effects. This does introduce some model 
dependence in the signal determined by subtracting the background 
from the total correlation to the extent of inaccuracies of the 
background. However, the robustness and characteristics of the signals obtained
by this method imply that these signals are reasonably accurate. One should 
note that in comparing with theoretical models (Section VI) we use total 
correlations in the model and data analysis to eliminate any model dependence 
due to the separation of signals and background.
                                                      
In this paper we present an analysis of charged particle pair correlations in
two dimensions --- $\Delta \phi$ and $\Delta \eta$ ---  based on 13.5 million
minimum bias Au + Au events observed in the STAR detector at $\sqrt{s_{NN}}$ =
200 GeV \cite{STARTPC}\footnote{$\Delta \phi = \phi_1 - \phi_2$ and
$\Delta \eta = \eta_1 - \eta_2$}. One should note that only the minimum bias
trigger was used.
                                                                       
The paper analyzes independently the correlations in each of the STAR
minimum bias centrality bins from 0 to 80\% \cite{minbias} and then makes
comparisons with quantitative models and draws conclusions. The data
analysis is very similar to that used in Ref. \cite{centralproduction}; the
differences will be discussed. Data cuts were applied to make track merging
effects, HBT correlations, and Coulomb effects negligible.
                                                                             
The analysis leads to a multi-term correlation function similar to that of
equations (3 + 4 + 5) of Section III C and E of Ref. \cite{centralproduction}.
This multi-term function fits the $\Delta \eta \Delta \phi$ distribution well.
It includes terms describing correlations known to be present: collective flow,
momentum and charge conservation, and instrumental effects (equation (1)
Section III C). The sum of these terms are considered background defined as
\bf Bk \rm in equation (2) (Section III A-C). What remains are correlations
which we assume to represent our signals; equations (3) and (4) Section III E.
Their robustness, characteristics and significance in the fits
are clearly consistent with their being signals in the unlike-sign charge pairs
(US) and like-sign charge pairs (LS). Without the signal terms the fits to our
data are highly rejected, but with the addition of the signal terms the data
are well fit with reasonable parameterizations of the signals such as discussed
in Ref. \cite{centralproduction}.
                                                                          
This paper is organized as follows:
                                                                         
Section II describes data utilized and method of data analysis. Section III
describes parameterization of the data. Section IV discusses the CI and CD
signals and has a comparison with other data. Section V discusses systematic,
parameter, and fit errors. Section VI contains a discussion and comparison
with models. Section VII contains Summary and Conclusions. The last section is
an Appendix A which contains parameters of the fits as a function of
centrality percentage.
                                                              
\section{Data Analysis}

\subsection{ Data Utilized}
The data reported here is from STAR events taken at RHIC during the 2004 
running period for Au + Au collisions at $\sqrt{s_{NN}}$ = 200 GeV. 
The data were taken using a minimum bias trigger with the full STAR magnetic 
field (0.5 Tesla).

The experimental arrangement is very similar to that described in 
Ref. \cite{centralproduction}. The data analysis and parameterization methods 
used previously for the most central (approximately 0-10\%) centrality region 
were employed with slight modification. The same parameterization was 
successfully used in each of the 9 centrality bins in this analysis. These 
parameters (see Section III and Appendix A) were independently fit in each of 
the nine centrality bins.

About half the data were taken with the magnetic field parallel to the beam
axis direction (z) and the other half in the reverse field direction in 
order to determine if directional biases are present. The procedure described
later in this subsection for our $\chi^2$ analyses demonstrated there was
no evidence of any difference in the data samples from the two field
directions: the $\chi^2$ distribution of the difference of the two field
directions in the $\Delta \phi$ and $\Delta \eta$ bins were consistent with a
normal distribution expected for the Degrees of Freedom (DOF). There was no 
evidence for directional biases.

The track reconstruction for each field direction was done using the same 
reconstruction program. Events used in the analysis were required to have at 
least 14 primary tracks lying inside $|\eta| < 0.5$. Tracks that we use are 
required to have at least 23 hits in the TPC (which for STAR eliminates split 
tracks), and have pseudo-rapidity, $\eta$, between -1 and 1. These are tracks 
that are consistent with the criteria that they are produced by a Au + Au 
interaction. The surviving events totaled 7.6 million for the forward
field and 5.9 million for the reverse field. The transverse momentum selection 
$0.8 < p_{t}  < 4.0$ GeV/c was then applied. The upper $p_t$ limit of 
\cite{centralproduction} was raised to 4 GeV/c since, as discussed in
\cite{PBM} (Section 1, last paragraph) jet contamination is negligible if you
 use minimum bias data without a jet trigger.

Each of the nine centrality bins was treated separately and fit separately.
In each centrality bin the event records were sorted according to the z
(collider axis) position of the primary vertex into ten 5 cm wide bins along
z from -25 cm to +25 cm relative to the center of the STAR TPC. This produced 
ten files for each sign of the magnetic field in each centrality bin. The 
events for the same $z$ bin, thus the same acceptance, were then merged to 
produce 20 files, one for each z vertex bin, for each sign of the magnetic 
field.

The files were analyzed in two-dimensional (2-D) histograms of the difference 
in $\eta$ ($\Delta \eta$), and the difference in $\phi$ ($\Delta  \phi$) for 
all the track pairs in each event. Each 2-D histogram for each centrality bin
had 72 $\Delta \phi$ bins ($5^\circ$) from $-180^\circ$ to $180^\circ$ and 
38 $\Delta \eta$ bins (0.1) from -1.9 to 1.9. The sign of the difference 
variable was chosen by labeling the positive charged track as the first of the 
pair for the unlike-sign charge pairs, and the larger $p_t$ track as the first 
for the like-sign charge pairs. Our labeling of the order of the tracks in a 
pair allows us to range over four $\Delta \phi$ - $\Delta \eta$ quadrants, and 
to investigate possible asymmetric systematic errors due to geometry, magnetic 
field direction, behavior of opposite charge tracks, and $p_t$ dependence. 
Our consistently satisfactory results for our extensive $\chi^2$ tests of the
data for these quadrants revealed no evidence for such effects.

Then we compared the $\Delta \phi$ - $\Delta \eta$ data for the two field
directions on a bin by bin basis. In the reverse field data, we reversed the
track curvature due to the change in the field direction, and changed the sign 
of the z axis making the magnetic field be in the same direction as the 
positive z direction. This is done by reflecting along the z axis, and 
simultaneously reflecting along the y axis. In the two dimensional 
$\Delta \phi$ - $\Delta \eta$ space this transformation is equivalent to a 
reflection in $\Delta \phi$ and $\Delta \eta$. For each pair we changed the 
sign of its $\Delta \phi$ and $\Delta \eta$ in the reverse field data. We then 
calculated a $\chi^2$ based on the difference between the forward field and the
reverse field, summing over the $\Delta \phi$ - $\Delta \eta$ histograms 
divided by the errors added in quadrature. The $\chi^2$ distribution of the
difference of the two field directions in the $\Delta \phi$ and $\Delta \eta$ 
bins was consistent with a normal distribution for random 
fluctuations of a statistical nature.  Therefore considering the above clear 
justification we added the data for the two field directions in each 
centrality bin.

We compared the central z vertex bins with the outer z vertex bins in each
individual centrality bin. We found no evidence of differences. Therefore we
added the files for those 10 bins in each centrality bin.

\subsection{Analysis Method}
Separate $\Delta \phi$ - $\Delta \eta$ histograms were made in each centrality
bin for the two basic pair types unlike-sign
charge pairs (US) and like-sign charge pairs (LS) from the same-event-pairs, 
since their characteristics were different. Both histograms are needed later
to determine the CD and the CI correlations. Similar histograms were made 
with each track paired with tracks from a different event (mixed-event-pairs), 
adjacent in time, from the same z vertex bin and thus the same acceptance. 
This allows use of the technique of dividing the histograms of the 
same-events-pairs by the histograms of the mixed-events-pairs which 
strongly suppresses instrumental effects such as acceptance etc., but leaves 
small residual effects. 

The resulting 2-D  total correlation function is defined by
\begin{eqnarray*} 
 C(\Delta \phi, \Delta \eta)= S(\Delta \phi, \Delta \eta)/M(\Delta \phi, \Delta \eta)
\end{eqnarray*}
 where $S(\Delta \phi, \Delta \eta)$ is the number of pairs at the 
corresponding values of $\Delta \phi \Delta \eta$ coming from the same event 
after having summed over all the events. $M(\Delta \phi, \Delta \eta)$ is the 
number of pairs at the corresponding values of $\Delta \phi \Delta \eta$ coming
from the mixed events after we have summed over all our mixed events. 
$C(\Delta \phi, \Delta \eta)$ is constructed separately for US and LS pairs and
each is normalized to a mean of 1.

The symmetries in the data allowed us to fold all four 
$\Delta \phi$ - $\Delta \eta$ quadrants into the one quadrant where both 
$\Delta \phi$ and $\Delta \eta$ were positive. After the cuts described later
in this subsection, we compared the unfolded bins to the folded average for 
unlike-sign charge pairs and like-sign charge pairs separately. 
The folded and unfolded distributions were statistically consistent.
We searched in a number of ways to find asymmetries in the data via 
extensive $\chi^2$ analyses and observation of fit behavior; none of any 
significance were found. By folding four quadrants into one we quadrupled the 
statistics in each bin analyzed. 

Henceforth the folded data will be used for our fits which increases our 
statistics per bin by a factor of 4. We used the same method of $\chi^2$ 
as described in Ref. \cite{centralproduction} Section II C to eliminate 
bins which exhibited non-negligible track merging effects. At 
small $\Delta \phi$ and $\Delta \eta$ (i.e. small space angles) track 
merging effects occur. To determine the cuts needed to reduce 
these effects to a negligible level, we varied small $\Delta \phi$ 
and small $\Delta \eta$ cuts. Simultaneously the $\chi^2$ of an 
approximate fit to the data using equations (2) + (3) + (4) (see Sections III)
was studied as a function of the bins included in the fit. With larger cuts the
$\chi^2$ behaved properly until one or more of the bins affected by merging was
included in the fit. This caused a huge increase in $\chi^2$, revealing that 
those bin(s) were distorted. We confirmed by visual inspection that
track merging clearly became important in the bins eliminated which caused a 
substantial reduction in track recognition efficiency. The resultant 
cuts also made the HBT and the Coulomb effects negligible. The required 
removal of bins was made by the following cuts: 

For the unlike-sign charge pairs (US) 0.0 $< \Delta \eta <$ 0.1 and $0^\circ$
$< \Delta \phi <$  $20^\circ$, and  0.1 $< \Delta \eta <$ 0.2 and $0^\circ$
$< \Delta \phi <$  $10^\circ$ were eliminated.

For the like-sign charge pairs (LS) 0.0 $< \Delta \eta <$ 0.2 and $0^\circ$
$< \Delta \phi <$  $10^\circ$ were eliminated.
 
The track topology differs for US and LS pairs due to their different 
curvatures in the magnetic field. The two tracks in the unlike-sign
charge pairs curve in opposite directions while for the like-sign charge
pairs the two tracks curve in the same direction. This makes the merging
characteristics different and requires the cuts to be different in order to
make the track merging effects negligible.
 
The fits to the data were made over the whole $\Delta \phi$ and $\Delta \eta$
ranges except for the above cuts. The data for $|\Delta \eta| >$ 1.5 were cut 
out of the fits, since statistics were low and variations in efficiency are 
large. However the fits create an extrapolation for small $\Delta \eta$ and 
$\Delta \phi$ bins where we have cut and thus correct the fits for the loss of 
the cut-out small angle bins.

\section{Parameterization of data.} 
We wanted to obtain a set of functions which will fit the data well and are 
interpretable to the extent practical. We utilized parameterizations
representing known, expected physics, or attributable to instrumentation 
(or other non relevant to this analysis) effects. Any remaining terms required 
to obtain good fits to the data can be considered as signals of new physical 
effects. Thus signal $\equiv$ data - (known and expected) effects. For the 
three known effects (elliptic flow, residual instrumental effects, momentum and
charge conservation) appropriate terms were parameterized. We then found 
parameters for the signal terms which are necessary in order to achieve a good 
fit to our high statistical precision data.

\subsection{Parameters Related to Elliptic Flow}

The parameters related to elliptic flow \cite{flow} were represented by the 
usual large term $2 v_2^2\rm\cos(2 \Delta \phi)$. We also needed a 
$\rm\cos(4 \Delta \phi)$ term in order to obtain good $\chi^2$ fits in our
intermediate $p_t$ range.

\subsection{Instrumental Effects}

There is a $\phi$ independent effect which we attribute to losses in the larger
$\eta$ tracking in the TPC and perhaps part of a long range correlation which 
is of no interest to this analysis and cannot be analyzed in detail because of 
the smallness of the effect. We utilized mixed-event-pairs with a similar 
$z$-vertex to take into account these effects. Imperfections in this procedure 
leave a small depression near larger $\Delta \eta$ to be represented in the 
fit by a Gaussian term labeled ``etabump amp'' and ``etabump width''. The 
width of this effect should be independent of the charge of the tracks, so we 
constrained it to be the same for like and unlike-sign charge pairs to improve 
the fit stability. We found that choosing a fixed center of 1.25 and a fixed 
width $\sigma_E$ of 1.57 for all centrality bins was adequate. However the 
amplitude was allowed to vary. Thus the functional form for this effect 
(which was treated as a background) is:
 
\begin{equation} \rm Etabump \it (\Delta \eta) = E e^{-(\Delta \eta - 1.25)^2/2\sigma_E^2} \end{equation} 
                           
The residual sector effects due to the lack of readout in the boundary regions
(gaps between the 12 readout sectors) clearly needed corrections for the 
2001 data utilized in Ref. \cite{centralproduction}. However in this analysis 
the division by mixed pairs was sufficient to make them negligible. This was 
due to the fact that in the 2004 run, the data were corrected for the space 
charge event by event. We had attributed the need for residual sector 
corrections in the 2001 run to space charge effects \cite{centralproduction}. 

\subsection{Correlations associated with Momentum and Charge Conservation}
It is important to ensure that momentum and charge conservation correlation 
requirements are satisfied. For random emission of single particles with 
transverse momentum conservation globally imposed, a negative 
$\rm\cos$($\Delta \phi$) term alone can represent this effect since random
emission of single particles results in no correlations between them. It has
been shown \cite {borghini} that the $\rm\cos$($\Delta \phi$) term alone is
correct for conserving transverse momentum when there are no other
correlations present.  However, the complex correlations that occur at RHIC 
result from multiple sources which are presently not understood. It was not 
possible to fit our data with the $\rm\cos$($\Delta \phi$) term alone. Fits 
were rejected by  $\sim40\sigma$ or greater for both the unlike-sign charge 
pairs and like-sign charge pairs in the 0-50\% centrality region. This was not 
surprising since random emission of single particles with transverse momentum 
conservation would not lead to the particle correlations observed at RHIC. 
Therefore we suspected that a more complete description of momentum and 
charge conservation was required. No one has succeeded in solving this complex 
problem in closed form even in the theoretical case where you observe all 
particles. It appears that a solution to this problem would require a knowledge
of all the correlations present in a particular analysis and this is not 
attainable in practice. Hence a reasonable approach was to try to solve it for 
the tracks we are observing in order to obtain a good fit. For the two 
variables we have, $\Delta \phi$ and $\Delta \eta$, we used Fourier analysis 
and polynomial expansion respectively.
 
\begin{figure*}[ht] \centerline{\includegraphics[width=0.800\textwidth]
{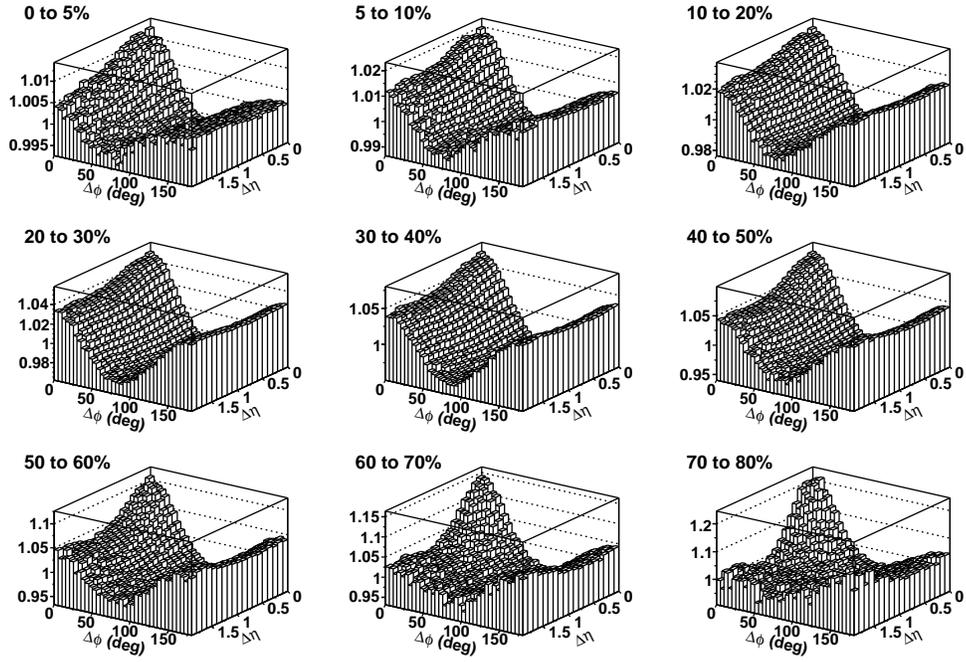}} \caption[]{The folded - after - cuts unlike-sign
charge pairs (US) correlation data vs. centrality.}
\label{figure1}
\end{figure*}
 
\begin{figure*}[ht] \centerline{\includegraphics[width=0.800\textwidth]
{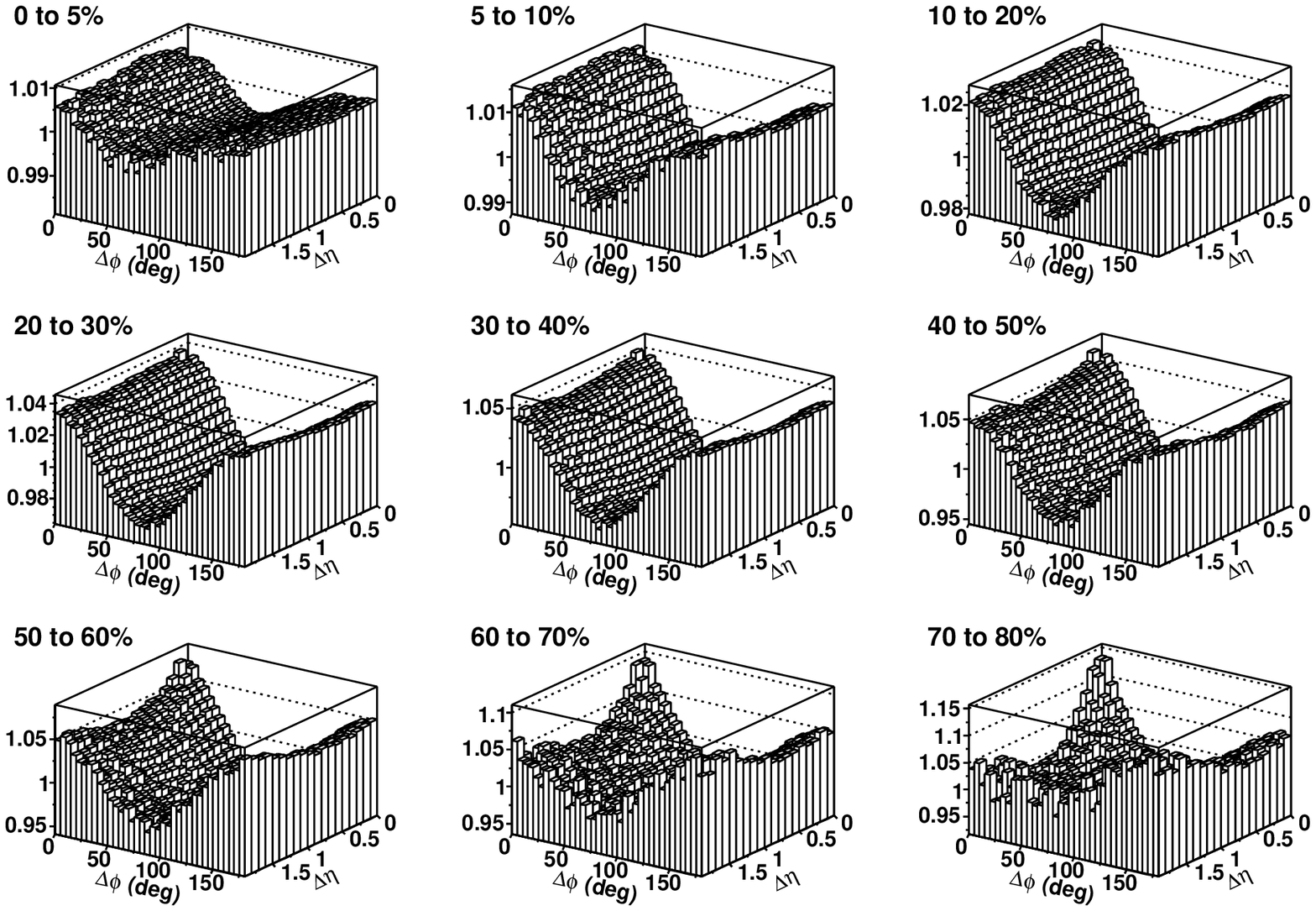}} \caption[]{The folded - after - cuts like-sign
charge pairs (LS) correlation data vs. centrality.}
\label{figure2}
\end{figure*}
                                                     
\begin{figure*}[ht] \centerline{\includegraphics[width=0.800\textwidth]
{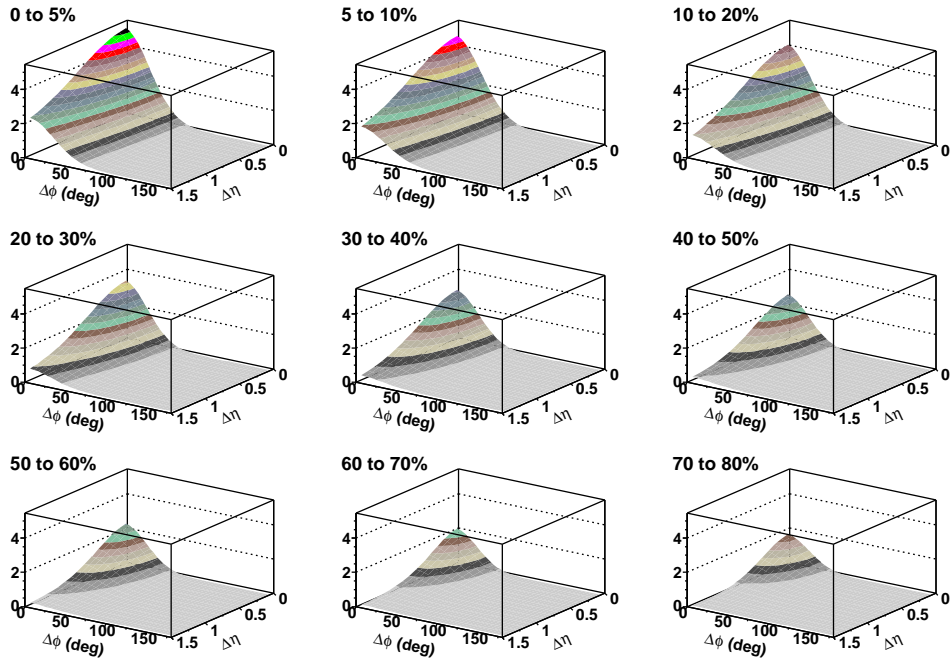}} \caption[]{``(Color online)'' The signals within the prior
FIG. 1-2 US (equation 3) and LS (equation 4) are the basic correlation data
building blocks from which the two physically significant correlation signals
the Charge Independent (CI = US + LS) signal and the Charge Dependent
(CD = US - LS) signal are built (Section IV). The signals are obtained by
subtracting the background (\bf Bk \rm = equation 2) from the observed total
correlation. In order to compare different centrality bins or different
experiments one must remove the dilution factor of 1/multiplicity of the
observed correlation signals caused primarily by the quadratic increase of
pair combinations. This is accomplished by multiplying the correlation by the
multiplicity given in the last column of Table I and is defined as multiplicity
scaling or multiplicity scaled. This figure displays a 2-D perspective plot of
the multiplicity scaled unlike-sign charge signals (background subtracted)
given in equation 3 multiplied by the multiplicity (multiplicity scaled)
plotted as a function of $\Delta \phi$ and $\Delta \eta$ vs. centrality.}
\label{figure3}
\end{figure*}
                                                                      
\begin{figure*}[ht] \centerline{\includegraphics[width=0.800\textwidth]
{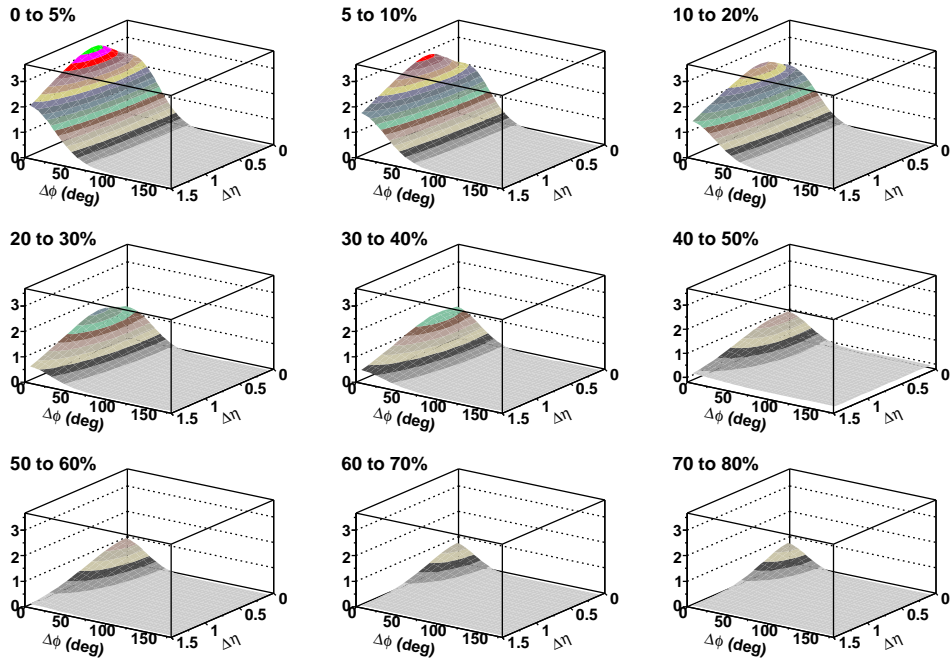}} \caption[]{``(Color online)'' a 2-D perspective plot of the
multiplicity scaled like-sign charge signals (background subtracted) given in
equation 4 vs. centrality plotted as a function of $\Delta \phi$ and
$\Delta \eta$.}
\label{figure4}
\end{figure*}
                                                                         
\begin{figure*}[ht] \centerline{\includegraphics[width=0.800\textwidth]
{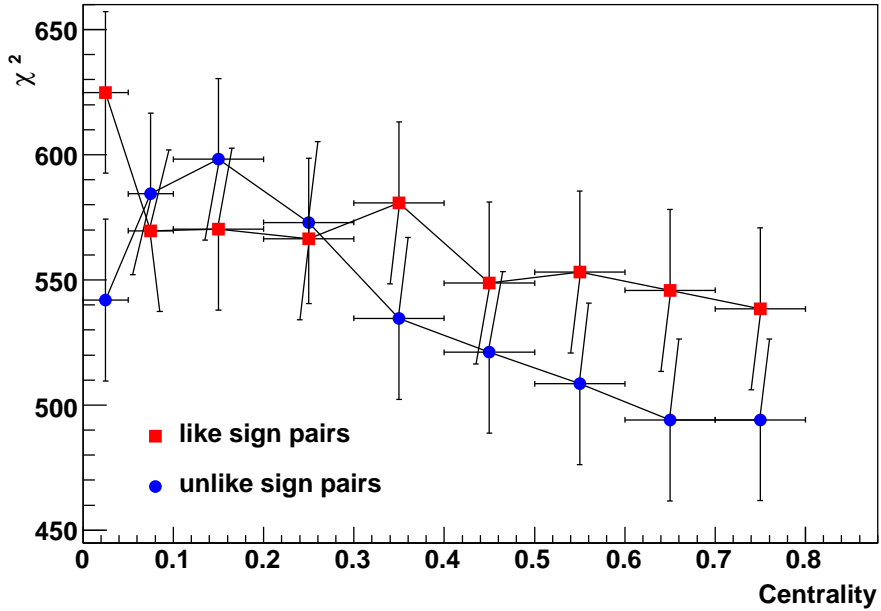}} \caption[]{``(Color online)'' The $\chi^2$ for both US and LS
vs. centrality using DOF as 521 and a $1\sigma$ degradation of the fit as equal
to a change of $\chi^2$ of 32 compared to 521 results in all fits being
consistent with $3\sigma$ or less. Some details relevant to the above procedure
are given in Section III D, and Section V.}
\label{figure5}
\end{figure*}
                                                
\begin{figure*}[ht] \centerline{\includegraphics[width=0.800\textwidth]
{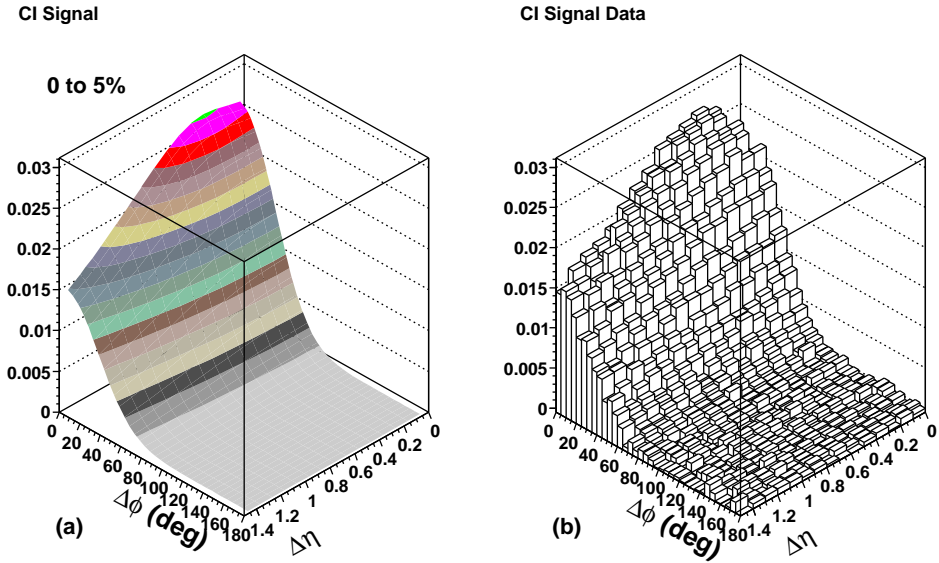}} \caption[]{``(Color online)'' a) 2-D perspective
plot fit to the CI signal (equation 3 + 4) in the 0-5\% centrality bin
(most central). Note that $\Delta \eta$ is elongated by a factor $\sim$3
compared to HIJING or Pythia jets, which agree with Fig. 7.
                                                                            
b) The CI signal data that were used in the fit.}
\label{figure6}
\end{figure*}
                                            
\begin{figure*}[ht] \centerline{\includegraphics[width=0.800\textwidth]
{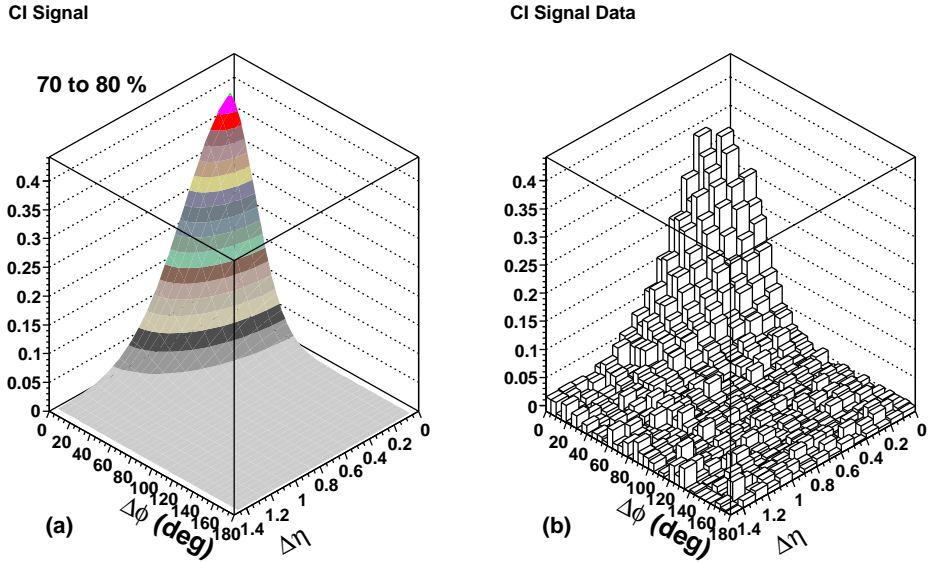}} \caption[]{``(Color online)'' a) 2-D perspective
plot fit to the CI signal in the 70-80\% centrality bin (most peripheral).
Note the large $\Delta \eta$ elongation has disappeared and we have approximate
jet-like symmetry (as seen in HIJING or Pythia jets).
                                                                        
b) The CI signal data that were used in the fit.}
\label{figure7}
\end{figure*}
                   
\begin{figure*}[ht] \centerline{\includegraphics[width=0.800\textwidth]
{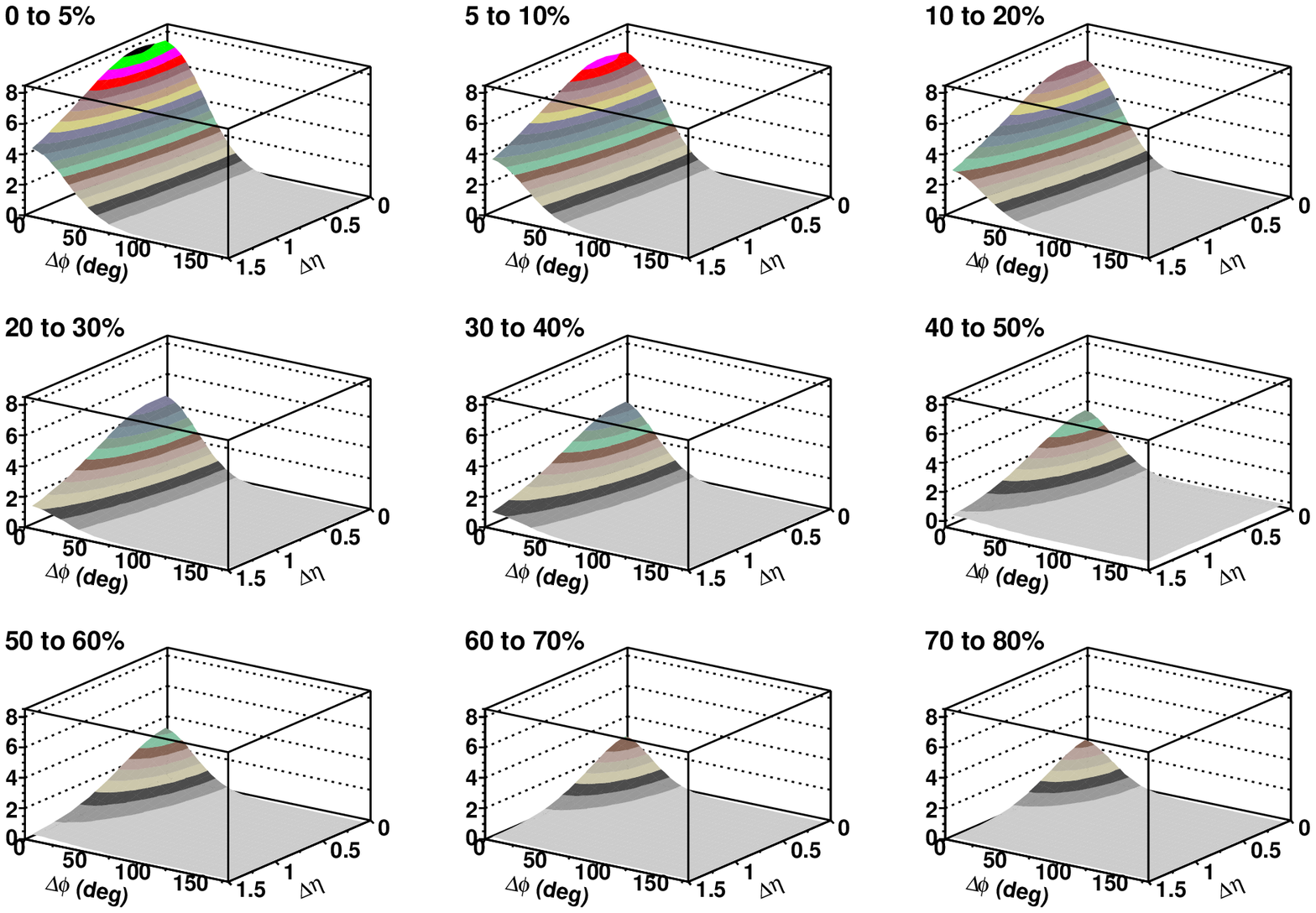}} \caption[]{``(Color online)'' The 2-D fits for the CI signal
multiplied by the multiplicity as a function of centrality. The CI displays
the average structure of the correlated emitting sources at kinetic freeze-out
(see Section IV).}
\label{figure8}
\end{figure*}

Assuming that the $\rm\cos$($\Delta \phi$) term for random single particle 
emission was the first term in a Fourier expansion of odd terms, a second term 
$\rm\cos$(3$\Delta \phi$) was added and found to account for almost all of the 
$\sim40\sigma$ rejection. Based on the residual analysis we concluded the 
remaining few \% required $(\Delta \eta)^2$ dependent terms for its removal in
order to obtain good fits. Therefore we multiplied terms of the type 
$\rm\cos$($\Delta \phi$) and $\rm\cos$(3$\Delta \phi$) by a $(\Delta \eta)^2$ 
term which when added reduced the remaining rejection and led to 
good fits.

In addition we found that we needed a  $(\Delta \eta)^2$ 
term in our background. Such a term was added to the background fit parameters 
since it is probably part of a long range correlation not relevant to this 
analysis.

If we take the sum of the terms described in subsections A, B and C
 we obtain the following for our background \bf Bk\rm.

 \begin{eqnarray}
\bf Bk \it = \rm (Known \it + \rm Expected)Effects \it = \nonumber \\
B_{00} + B_{02}(\Delta \eta)^2 + B_{10}\cos\Delta \phi 
+ B_{12}(\Delta \eta)^2\cos\Delta \phi \nonumber \\ 
 + 2 v_2^2\cos (2 \Delta \phi) + B_{30}\cos(3\Delta \phi) 
+ B_{32}(\Delta \eta)^2\cos(3\Delta \phi) \nonumber \\
+ B_{40}\cos(4\Delta \phi) + \rm Etabump \it (\Delta \eta) \nonumber \\
        \end{eqnarray}

\subsection{Fitting with Bk}

 We used the well known result \cite{probability} that for a large number of
degrees of freedom (DOF), where the number of parameters is a small fraction of
DOF and the statistics are high, the $\chi^2$ distribution is normally
distributed about the DOF. The significance of the fit decreases by $1\sigma$
whenever the value of $\chi^2$  increases by $\sqrt{2(DOF)}$ which for our 
521 DOF is equal to 32.

If we fit the functional form of the background \bf Bk \rm (equation (2)) to 
US in each centrality bin the fits are rejected by $\sim50\sigma$ or greater. 
If we fit LS to the background \bf Bk \rm in each centrality bin the fits are 
rejected by $\sim17\sigma$ or greater.

\subsection{Signal Terms and Multiplicity Scaling}

Many signal terms in physics are Gaussian-like. We therefore tried fitting the 
signal data using two dimensional (2-D) Gaussian or approximate Gaussian
parametric forms which successfully parameterized our signals in the
previous central production analyses \cite{centralproduction}. The values of the
signal and background (\bf Bk\rm) parameters in each centrality bin were 
independently fit for that particular centrality bin. 

The physical characteristics of elliptic flow have been extensively 
investigated. They are reasonably understood and are essentially 
charge independent. Therefore the same large flow $v_2$ term was used in 
US and LS fits and its value determined as part of our best fit to 
the data set. The US in each centrality bin shown in Fig 1 were well 
fit ($3\sigma$), by adding to \bf Bk \rm an additional 2-D approximate 
Gaussian in $\Delta \eta$ and $\Delta \phi$ given by:

\begin{equation} \rm Unlike\>-\>sign\> Signal \it  =  A_u e^{-((\Delta 
\phi)^2/2\sigma_\phi^2 +(\Delta \eta)^2/2\sigma_\eta^2 -f (\Delta \eta)^4)}
\end{equation}

Considering the enormous improvement in fit quality afforded by the addition
of this signal term, we conclude that this function in  equation (3) provides a
compact analytic description of the signal component of the unlike-sign charged
pairs correlation data. The fit was improved by the addition of a term 
dependent on $(\Delta \eta)^4$ in the exponent (called ``fourth''). This was
previously found to be the case in Ref. \cite{centralproduction}. 

Our normalization is such that the correlation is normalized to a mean of 1. 
When one compares different centralities the signal term is proportional to
the number of correlated signal particles within our cuts. This is so since 
when two particles form a correlated pair they are in general not correlated 
with the remaining signal particles. Thus the pool of correlated particles
is reduced by two every time a pair of correlated particles is picked. Hence
the signal term is proportional to the number of correlated signal pairs which
is 1/2 the number of signal particles within our cuts. On the other hand 
the number of entries to the correlation calculation grows as the number of 
particles squared. This quadratic increase dilutes the signal by a large factor
of 1/(particles). If we multiply the signal by efficiency corrected 
multiplicity (column 3 table I) we cancel this dilution. For all signal 
comparisons as a function of centrality and comparisons with other experiments
we utilize signal X multiplicity or equivalently multiplicity scaling or 
multiplicity scaled. The multiplicity used is the efficiency corrected 
multiplicity. The observed average multiplicity in the TPC and the efficiency 
corrected average multiplicity for each centrality is given in table I. The 
fits for the multiplicity scaled US signal data as a function of centrality for 
the folded - after - cuts data are shown in Fig. 3. 

\begin{center}
\begin{table}
\begin{tabular}{ c r r}\hline\hline
Centrality  & Average Multiplicity TPC & Corrected \\ \hline
0 to 5\% &  216 & 292\\ \hline
5 to 10\% &  180 & 237\\ \hline
10 to 20\% &  140 & 176\\ \hline
20 to 30\% &  98 & 120\\ \hline
30 to 40\% &  65.4 & 78\\ \hline
40 to 50\% &  41.5 & 48.2\\ \hline
50 to 60\% &  24.4 & 27.8\\ \hline
60 to 70\% &  13.1 & 14.6\\ \hline
70 to 80\% &  6.2 & 6.9\\ \hline
\end{tabular}
\caption[]{The average number of particles detected per event with 
$0.8 < p_t < 4.0$ and $|\eta| <1.0$ for the 9 centrality bins are given for 
the observed TPC particles (middle column), and the efficiency corrected 
multiplicity (last column).}
\end{table}
\end{center}

The LS data which also could not be fit by \bf Bk \rm 
alone, were well described ($3\sigma$) when we added (see Fig. 2) a positive 
2-D Gaussian and a small negative 2-D Gaussian dip given by: 
 
\begin{eqnarray} \rm Like\>-\>sign\> Signal \it  = A_l e^{-((\Delta \phi)^2/2\sigma_{\phi l}^2 +(\Delta \eta)^2/2\sigma_{\eta l}^2)} \nonumber \\
+ A_d e^{-((\Delta \phi)^2/2\sigma_{\phi d}^2 
+(\Delta \eta)^2/2\sigma_{\eta d}^2)}\end{eqnarray}

The significance of the small dip is $\sim20\sigma$ in the 0-30\% centrality
region. This dip is a physical effect not due to track merging. Since gluons 
prefer to emit US (pairs) and suppress emission of LS (pairs) in the same 
phase space region we attribute the dip in the LS as likely due to this 
suppression. The US (pairs) do not have a dip as expected.

Therefore, we conclude that equation (4) provides an efficient description of
the signal component of the LS data. The large signal is referred to as 
``lump'' in the LS and is a 2-D Gaussian centered at the origin. It is 
accompanied by a small narrower 2-D Gaussian ``dip'' (also centered at the 
origin) subtracted from it. Fig. 4 shows a 2-D $\Delta \phi \Delta \eta$ 
perspective plot of the multiplicity scaled LS signals (background subtracted) 
as a function of centrality.

All the above fits are consistent with significance of $3\sigma$.
Note that both US and LS signal fits show a $\Delta \eta$ elongation in the 
central region such that the corresponding angle is much larger than the
$\Delta \phi$ angle. These $\Delta \eta$ elongations reduce as centrality
decreases and virtually disappear in the peripheral region leading to 
approximately jet-like symmetry.

Appendix A contains the plots of the fitted parameters as a function of
centrality which produce the US and the LS fits accompanied by explanatory
material. The signals are to some degree model dependent because of the 
background (\bf Bk\rm) subtraction, but are robust and represent the 
characteristics of the underlying structures. Fig. 5 shows the dependence of
$\chi^2$ as a function of centrality bin for the fits of both the US and LS.
A larger value of $\chi^2$ above 521 means we have exceeded the DOF. The
difference of this $\chi^2$ divided by 32 gives the number of $\sigma$'s by
which the value of $\chi^2$ departs (e.g. is increased) compared to an ideal 
fit ($\chi^2$ = DOF). All the parameters fit together, signal plus background, 
represent a model independent measure of the correlation function, which can 
be theoretically analyzed in any way chosen.

The complex multi-dimensional $\chi^2$ surface makes the $\chi^2$ change non 
linearly with the number of error ranges ($1\sigma$) shown on the plots in 
Appendix A. Therefore, in order to determine the significance of a parameter
or group of parameters one must fit without them, and determine by how many
$\sigma$ the fit has worsened. Then one uses the normal distribution curve to
determine the significance of the omitted parameter(s).

\section{CI and CD Signals}

\subsection{Charge Independent (CI) Signals}
 
We obtain our signals by subtracting the background (\bf Bk\rm) given in
equation (2) from the total US and LS correlations. If we add the US signal
which has background subtracted to the LS signal with its 
background subtracted, we obtain the CI signal = US signal + LS signal. 
The CI signal fit and the data with background subtracted in the 
most central 0-5\% bin that was fitted are shown as 2-D perspective plots in 
Fig. 6a-b. The CI signal displays the average structure of the correlated 
emitting sources at kinetic freeze-out. One should note the large elongation of
the corresponding $\Delta \eta$ in the 0-5\% bin was previously observed in 
Ref. \cite{centralproduction}. The CI signal fit and the corresponding data in 
the most peripheral (70-80\%) bin that were fit are shown as 2-D perspective 
plots in Fig. 7a-b. We note the large $\Delta \eta$ elongation has disappeared 
and we have approximate jet-like symmetry.

In order to compare signals in different centrality bins or to compare with
different analyses we must use multiplicity scaling. In Fig. 8 we show
the multiplicity scaled CI signals for all the nine centrality bins. In the 
0-20\% region the large $\Delta \eta$ elongation mostly persists. In the 
20-60\% region this $\Delta \eta$ elongation gradually decreases with the 
decrease in centrality. In the two most peripheral bins the elongation is gone.

In Fig. 9 we compare the central region CI signal multiplied by multiplicity
for the 
present 2004 minimum bias Au + Au analysis with the previous 2001 central
trigger analysis. They appear quite similar and considering the differences
 in the analyses and the errors they agree well.

The $\chi^2$ for the fits to each CI centrality bin are consistent with a
significance of $3\sigma$ or better. Fig. 10 shows the multiplicity scaled
peak CI signal vs. centrality, and the integral of the multiplicity scaled CI
signal vs. centrality. Both have their maximum at the most central bin, and
decrease continuously by a factor of approximately 3 (amplitude) and 8 
(integral) as one proceeds to their minimum values in the most peripheral bin. 
A comparison with the prior central trigger analysis \cite{centralproduction} 
shows agreement within the errors in the overlap region. Also shown is a 
comparison with the results of a $\sqrt{s_{NN}}$ = 130 GeV analysis with a 
lower $p_t$ cut of 0.15 GeV/c \cite{aya}. We attribute the observed differences,
especially the dip at the most central CI bins, to the preponderance of low 
energy particles in that analysis.

Fig. 11 shows the $\Delta \eta$ width as a function of centrality for the
different analyses, while Fig. 12 shows the $\Delta \phi$ width for the same
centralities. The present analysis and the central trigger prior analysis agree
well. However in the 130 GeV analysis the $\Delta \phi$ width is decreasing
with increasing centrality, while in the 200 GeV analysis the $\Delta \phi$ 
width is essentially flat within the errors. However the trend shows a modest 
increase with centrality. We attribute the difference in behavior to the 
preponderance of low energy particles in the 130 GeV analysis. 

\subsection{Charge Dependent (CD) Signal} 
If we subtract the total LS correlation from the total US correlation we 
obtain the CD correlation. However, it is observed that the background 
\bf (Bk) \rm of the two terms are close enough in value within the errors to 
cancel each other in the subtraction. Thus the CD signal is essentially the 
same as the total CD correlation. The subtraction of the LS in the CD is 
considered equivalent to removal of the opposite sign charge pairs which are 
not from the same space time region \cite{centralproduction,balfun}. Therefore 
if no further interactions occurred after the opposite charge pairs 
hadronization the CD would represent the initial correlation of those opposite
sign charge pairs which are emitted from the same space and time region.

The CD signal and the corresponding data in the most central bin (0-5\%) are 
shown as 2-D perspective plots in Fig. 13a-b. The large $\Delta \eta$ 
elongation found in the CI is mostly gone in the CD most central bin, and we 
find close to jet-like symmetry. In the subtraction in forming the CD signals
the background terms nearly cancel out leaving small residual background 
terms. Therefore to take into account this systematic we use a procedure 
described in Section V.

The CD signal and the corresponding data in the most peripheral (70-80\%) bin 
that were fit are shown as 2-D perspective plots in Fig. 14a-b. Here we find 
virtually jet-like symmetry.

Fig. 15 shows the multiplicity scaled CD signal vs. centrality for all 
0-80\%  bins. All nine plots have the same approximately jet-like shape.

Fig. 16 compares the multiplicity scaled CD signal for the present 2004 minimum
bias central region (0-20\%) with the 2001 STAR/RHIC central trigger 
analysis. They agree well.

Fig. 17 shows the multiplicity scaled CD signal peak amplitude vs. centrality. 
The CD signal peak amplitude shows a decreasing trend towards the most 
peripheral bins. A comparison with the 2001 analysis (run 2) is quite good.

Fig. 18 shows the CD $\Delta \phi$ width vs. centrality. Within the errors it
is independent of centrality. The comparison with the 2001 central trigger 
analysis is in good agreement and the 130 GeV analysis is also 
approximately within the errors. 

Fig. 19 shows the CD $\Delta \eta$ width vs. centrality. Within the errors it
is independent of centrality. The comparison with the 2001 central trigger 
analysis is in good agreement and the 130 GeV analysis is also within 
the errors.

All the present analysis fits for the CD quantities as a function of
centrality are consistent with $3\sigma$ significance. The CD shape parameters 
do not change much with centrality (Fig. 18-19), and have close to jet-like 
behavior.

The CD at the time of formation (hadronization) is a measure of the 
initial average distribution of the correlation between the US particle pairs 
emitted from the same space time region. This initial correlation distribution 
is expected to be jet-like and very similar to Pythia \cite{pythia} and 
HIJING \cite{Wang} jets. If there were appreciable interaction with the fireball
medium before kinetic freeze-out the initial jet-like correlation observed 
would be changed. However the observed CD correlation is approximately 
consistent with Pythia and HIJING jet correlations at all centralities 
(see Fig. 18-19 and 22). Therefore we can conclude that the emitted pairs have 
little further interactions after hadronization. Thus surface or near surface 
hadronization and emission from the fireball both occur in the central region 
and all other centralities where there is appreciable particle density. In the 
most peripheral bins the particle density is low enough to allow undisturbed 
fragmentation and thus no change in the CD correlation. Thus the CD behavior 
is consistent with a surface emission model such as Ref. \cite{PBM}.

\section{Systematic, Parameter, and Fits Errors}
 
Systematic errors were minimized using cuts and corrections. The cuts
(see Section II B) were large enough to make contributions from track
merging, Coulomb, and HBT effects negligible. Systematic checks 
utilized $\chi^2$ analyses which verified that the analysis 
results did not depend on the magnetic field direction, the 
vertex z coordinate, or the folding procedures (see Section II ). By 
removing the tracks with $|\eta| > $1.0 we keep the systematic errors 
of the track angles below about $1^\circ$ \cite{backbegone}.

In Section IV of Ref. \cite{centralproduction} we referred to a simulation in
Ref. \cite{themodel} which estimated that the background resonance
contribution to the CD correlation cannot be more than 20\% (see appendix B
of Ref. \cite{centralproduction}). The conclusion was that the effect of these 
resonances was primarily to increase the amplitude of the CD. From the 
resonance calculations discussed in appendix B of Ref. \cite{centralproduction} 
the background resonance contribution to the US correlation was estimated to
be 5\%. Both the Parton Bubble model \cite{PBM} and HIJING \cite{Wang} only have 
the background resonance contribution coming from the soft beam jet 
fragmentation. For the present $p_t$ range model calculations \cite{PBM,Wang}
we estimated the resonance background contribution is less than 1\%. 
 
\begin{figure*}[ht] \centerline{\includegraphics[width=0.800\textwidth]
{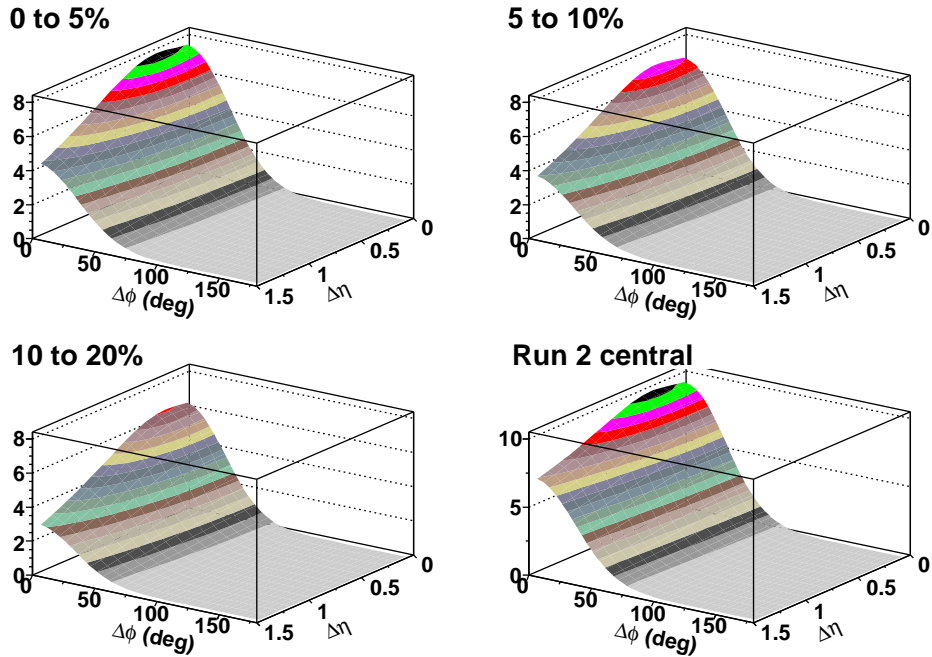}} \caption[]{``(Color online)'' A comparison of the
multiplicity scaled CI signal in the central region (0-20\%) for the present
2004 run and the earlier 2001 run (run 2) central trigger data.}
\label{figure9}
\end{figure*}
                                                            
\begin{figure*}[ht] \centerline{\includegraphics[width=0.800\textwidth]
{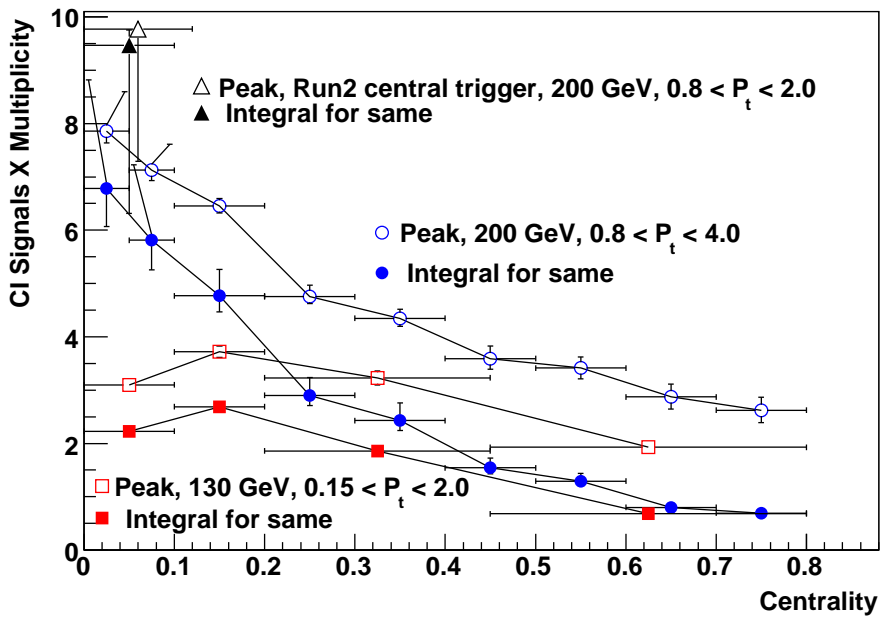}} \caption[]{``(Color online)'' The multiplicity scaled
peak CI signal  vs. centrality (blue open circular points). The blue solid
circular points show the integral of the multiplicity scaled CI signal vs.
centrality. The black triangular points show that the run 2 central trigger
results agree within the errors. The red open squares points show the
multiplicity scaled CI peak signal and the red solid square points show
the multiplicity scaled CI signal integral for the 130 GeV analysis vs.
centrality. The disagreement between the present analysis and the 130 GeV
analysis, especially the dip at the most central, is attributed to the
preponderance of low energy particles in the 130 GeV analysis.}
\label{figure10}
\end{figure*}
                                                                      
\begin{figure*}[ht] \centerline{\includegraphics[width=0.800\textwidth]
{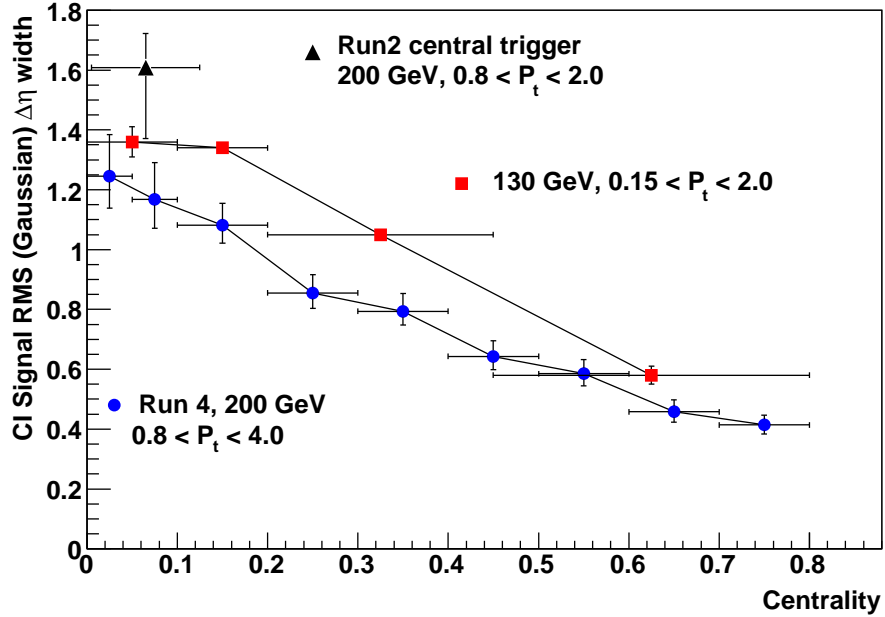}} \caption[]{``(Color online)'' The CI $\Delta \eta$ width vs.
centrality for the analyses indicated. RMS(Gaussian) means that a Gaussian
with the same RMS values as the actual fit was used to determine the width
values shown.  The differences between the present analysis and the 130 GeV
analysis are attributed to the preponderance of low energy particles in the
130 GeV analysis.}
\label{figure11}
\end{figure*}
                                                                
\begin{figure*}[ht] \centerline{\includegraphics[width=0.800\textwidth]
{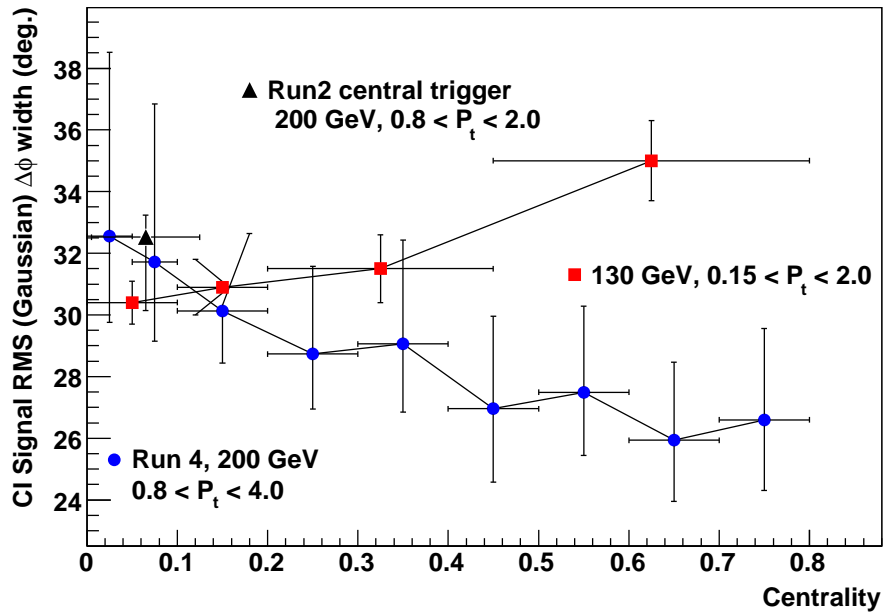}} \caption[]{``(Color online)'' The CI $\Delta \phi$ width vs.
centrality. The agreement of this analysis with Run 2 (central trigger) is
good. The disagreement between the present analysis and the 130 GeV
analysis is attributed to the preponderance of low energy particles in the 130
GeV analysis.}
\label{figure12}
\end{figure*}
                                              
\begin{figure*}[ht] \centerline{\includegraphics[width=0.800\textwidth]
{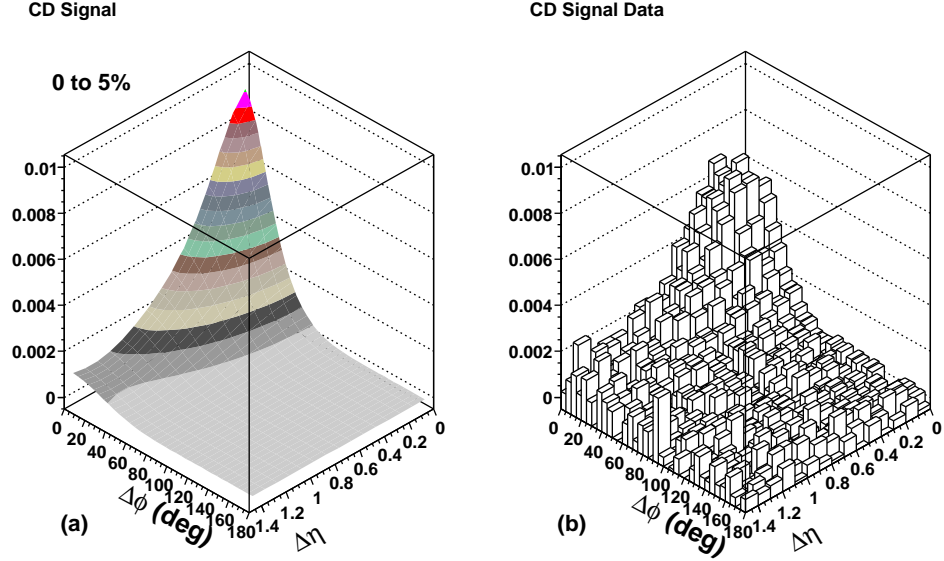}} \caption[]{``(Color online)'' a) 2-D perspective
plot fit to the CD signal in the 0-5\% centrality bin (most central).
Note that the large $\Delta \eta$ elongation found in the CI is mostly gone in
the CD most central bin, and we find close to jet-like symmetry.
                                                                         
b) The CD signal data that was used in the fit.}
\label{figure13}
\end{figure*}
                                                                         
\begin{figure*}[ht] \centerline{\includegraphics[width=0.800\textwidth]
{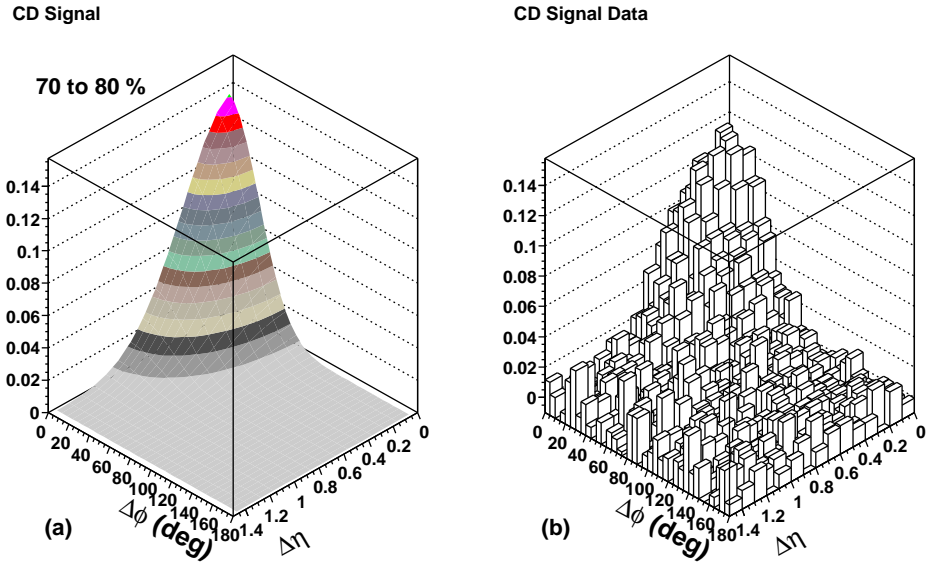}} \caption[]{``(Color online)'' a) 2-D perspective
plot fit to the CD signal in the 70-80\% centrality bin (most peripheral).
Note that like the CI the large $\Delta \eta$ elongation is not present, and
we find close to jet-like symmetry.
                                                                           
b) The CD signal data that was used in the fit.}
\label{figure14}
\end{figure*}
                                                                           
\begin{figure*}[ht] \centerline{\includegraphics[width=0.800\textwidth]
{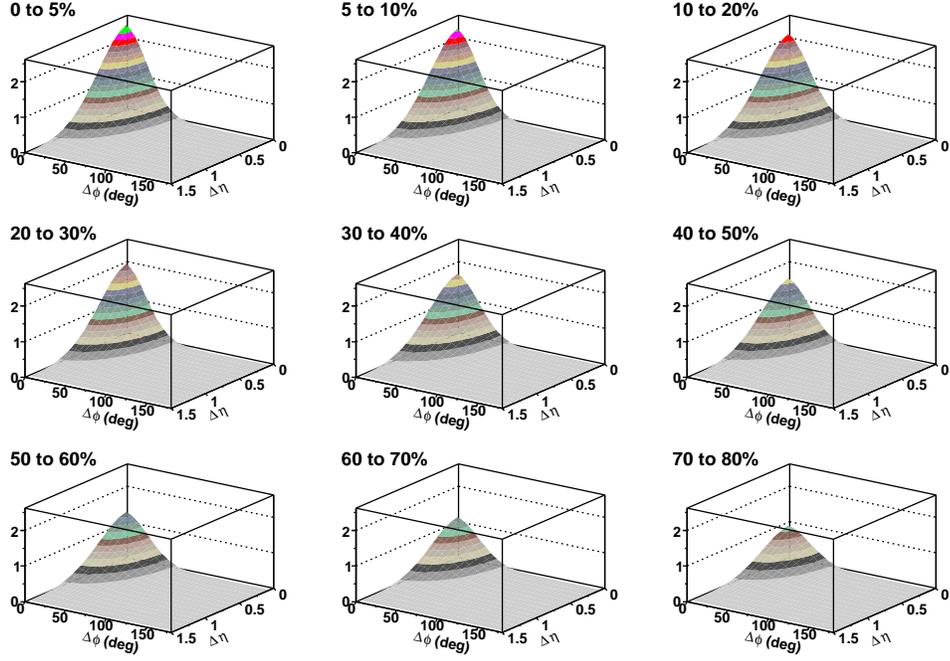}} \caption[]{``(Color online)'' The 2-D fits for the CD signal
multiplied by the multiplicity as function of centrality.}
\label{figure15}
\end{figure*}
                                                                         
\begin{figure*}[ht] \centerline{\includegraphics[width=0.800\textwidth]
{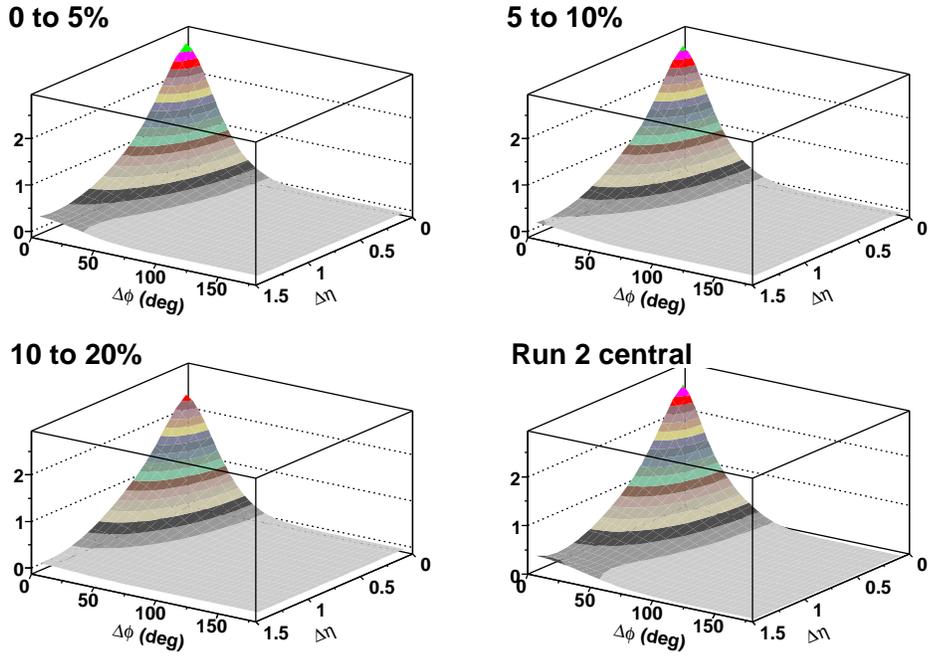}} \caption[]{``(Color online)'' Comparison of the central
multiplicity scaled CD signal with the previous 2001 (run 2) central triggered
analysis, which agree as shown in Fig. 17-19.}
\label{figure16}
\end{figure*}

\clearpage

\begin{figure*}[ht] \centerline{\includegraphics[width=0.800\textwidth]
{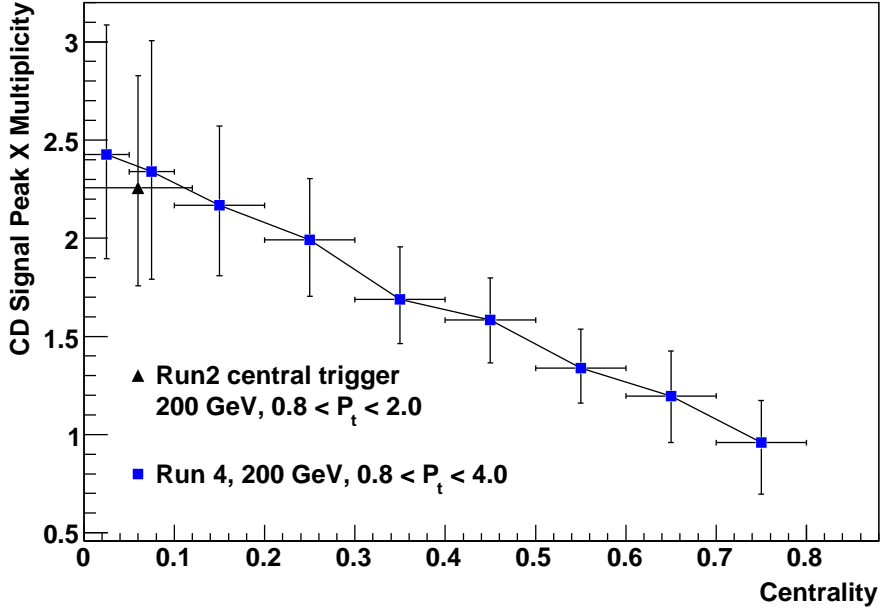}} \caption[]{``(Color online)'' The multiplicity scaled CD
signal peak vs. centrality and comparison to Run 2 which agrees well.}
\label{figure17}
\end{figure*}
                                                        
\begin{figure*}[ht] \centerline{\includegraphics[width=0.800\textwidth]
{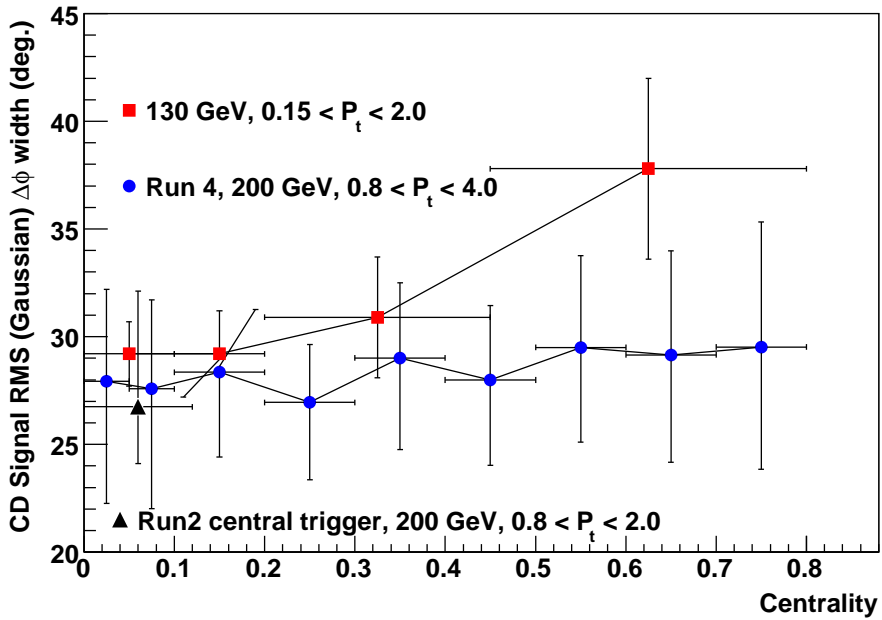}} \caption[]{``(Color online)'' The CD $\Delta \phi$ width
shown vs. centrality. The CD $\Delta \phi$ width is independent of centrality
for 200 GeV. The present and previous analyses agree within the errors.}
\label{figure18}
\end{figure*}
                                                                      
\begin{figure*}[ht] \centerline{\includegraphics[width=0.800\textwidth]
{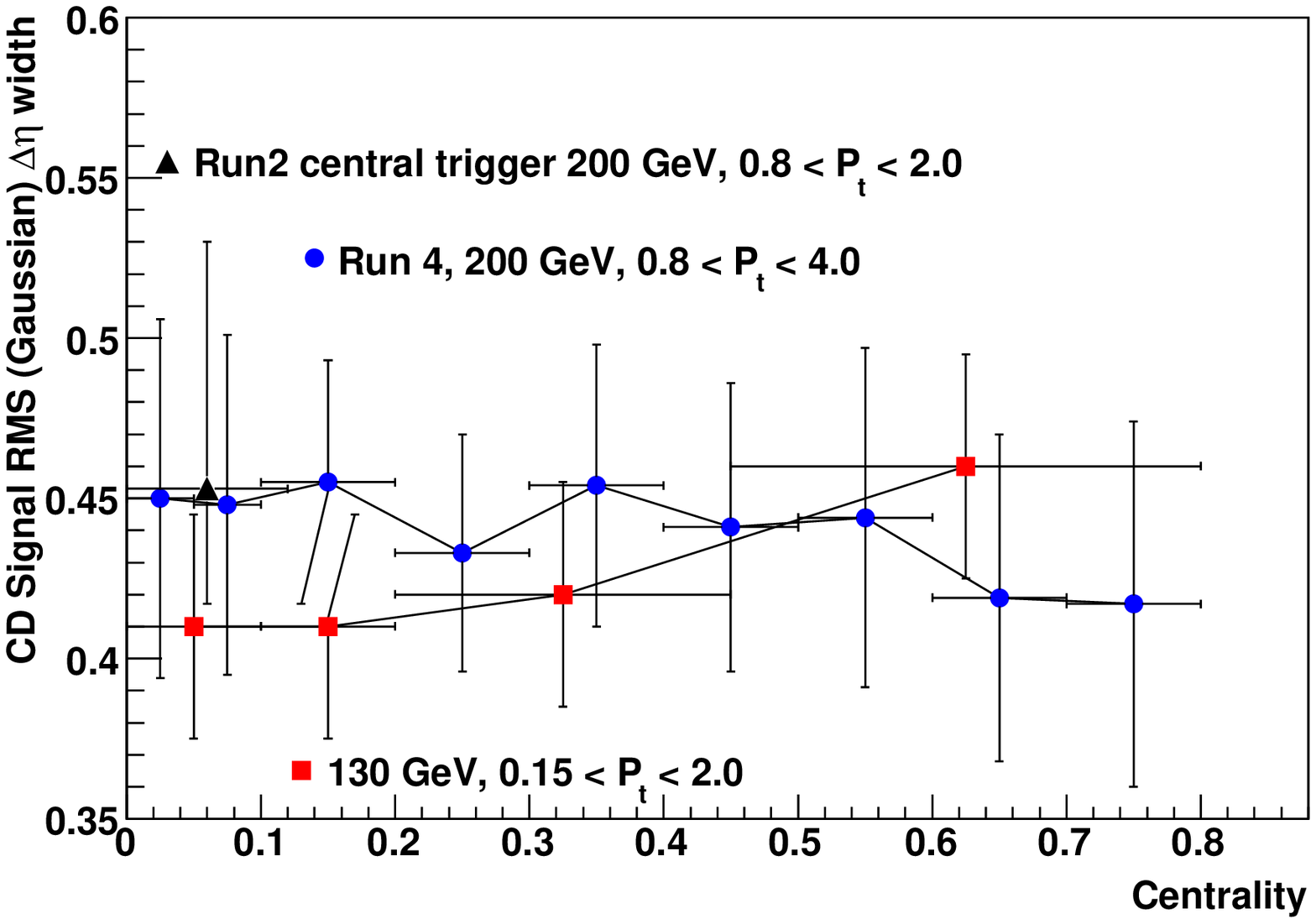}} \caption[]{``(Color online)'' The CD $\Delta \eta$ width vs.
centrality. The present and previous analyses agree within errors.}
\label{figure19}
\end{figure*}

As we shall see in the next section where we discuss model fits, the data can 
be well fit by assuming that background resonances are negligible.

In Section VI. Systematic Errors of Ref. \cite{centralproduction} we concluded
based on an extensive investigation for that paper that the increase in the
CI correlation $\Delta \eta$ width would be about 2\% at most. We estimate 
that in the central region of the present experiment where the $\Delta \eta$ 
width elongation is large this approximately 2\% estimate remains the same. As 
the CI elongation becomes smaller the error estimate is adequate since the CI, 
CD, US, and LS signals approach the same jet-like shape which has a negligible 
elongation. The multiplicity scaling errors are estimated to be about 2-3\%.
  
We have used the multi-parameter fitting procedure in the large DOF
region \cite{probability} that we employed in Ref. \cite{centralproduction}.
The reader is referred to Ref. \cite{probability} from which we quote: ``For
large n(DOF), the $\chi^2$ p.d.f. (probability density function) approaches a
Gaussian with mean = n and $\rm variance^2$ ($\sigma^2$) = 2n.'' For n$>$
50-100 this result has been considered applicable, and it remains applicable
and becomes more accurate as n increases toward infinity. It is important to
note that the above has been proven under the assumption that we do not know 
the underlying physics which is certainly the case for our analyses and almost
all heavy ion analyses at RHIC or elsewhere. In the US and the LS fits (with
521 DOF) an increase in $\chi^2$ of 32 reduces the fit significance 
by $1\sigma$ ($\sqrt{2DOF}$). The statistical significance
of any data analysis fit in this paper can be obtained by the following 
procedure: The number of $\sigma$'s of reduction of a fit significance =  
value of ($\chi^2$ - DOF)/32. The number of $\sigma$'s refer to the normal 
distribution curve and give the probability of the fit compared to an ideal
fit where $\chi^2$ = DOF. For large DOF (bins - parameters) fluctuations occur 
because of the many bins. When one fits the parameters, they will try and 
describe some of these fluctuations. Therefore we need to check whether the 
fluctuations in the data sample are large enough to significantly distort the 
parameter values. This has traditionally been done by using the confidence 
level tables vs. $\chi^2$ which allows a reasonable determination of the 
fluctuations due to binning. The approximation we have used 
is an accurate extension of the confidence level tables for the high ($>$ 100)
DOF region we are analyzing.
 
Let us consider the method of assigning systematic error ranges to the
parameters. Our objective is to obtain systematic error ranges which are not 
likely to be exceeded if future independent data samples taken under similar 
conditions are obtained by repetition of the experiment by STAR or others. We 
allow each parameter, one at a time, to be varied (increased and then 
decreased) in both directions while all the other parameters are free to 
readjust until the overall fit $\chi^2$ degrades in significance by $1\sigma$. 
This corresponds to a change of $\chi^2$ of 32 for both the US and LS fits. 
The $\chi^2$ surface has been observed, and $\chi^2$ increases very non 
linearly with small increases of the parameter beyond the error ranges. 
Thus this is a conservative method of assigning systematic errors to the
parameters determined.
 
The elliptic flow has a reasonably known underlying physics which is charge
independent. In the case of the CD elliptic flow effects have been observed to
cancel and cause negligible uncertainties. Therefore we used the same 
large elliptic flow $2 v_2^2\rm\cos$($2 \Delta \phi$) term for the US and LS 
fits. Thus our best fit has this same elliptic flow term in the US and LS 
which cancels out in the CD since it is obtained by subtraction of the LS from 
US correlation. 

The small difference of the $\rm\cos(4 \Delta \phi)$ terms in the CD 
correlation constitute a background that must be determined. There are two 
possibilities for determining the CD signal and background. In the first we 
defined the CD signal as the difference between the US and the LS signals. 
Thus the background is the difference between the US and LS backgrounds. This 
subtraction gives us amplitudes and widths determined by the $\chi^2$ surface 
of the US and LS fits. On the other hand we could directly subtract the LS 
correlation from the US correlation and obtain the CD correlation, which could 
also be fitted by a Gaussian plus background terms. These two procedures give 
a very similar result indicating that the small residual background terms are 
not a problem. We take the average of these two methods to define the CD signal
correlation amplitudes and widths. The systematic errors of these measurements
are determined by exploring the $\chi^2$ surface of the direct fit of the
CD signal correlation (method 2) plus the extracted measurements from the 
difference of the LS signal from the US signal using the $\chi^2$ surface of 
the US and the LS fits (method 1). 

Let us now address the errors due to contamination by including secondary
particles arising from weak decays and the interaction of anti-protons and
other particles in the beam pipe and material near the beam pipe. These
secondary particle backgrounds have been estimated to be about 10-15\%
 \cite{starback}. In this analysis we are concerned mainly with the angles of
the secondary particles  relative to the primaries that survive our high $p_t$
cut, not their identity or exact momentum magnitude. Our correlations almost
entirely depend on angular measurements of $\Delta \phi$ and $\Delta \eta$.

In the range $0.8 < p_t < 4.0$ GeV/c we have considered the behavior of
weakly decaying particles and other non-primary particles which could
satisfy our distance of closest approach to the primary vertex and $p_t$ cuts.
Because high $p_t$ secondary particles are focused in the same direction as
the primaries, only a fraction of these particles have sufficient change in
angle that would cause an appreciable error in our correlations. We utilized 
the methods developed in Ref. \cite{centralproduction} to estimate
that only 1/4 of the 10-15\% secondary particles satisfied our cuts and 
distance of closest approach. This led to our estimate that there is a 
systematic error of about 5\% due to secondary particles which is mostly an 
overall scale factor.
 
\section{Discussion and Model Fits.}
 
\subsection{Discussion}
 
The multiplicity scaled CI signal vs. centrality is shown in Fig. 8. as a 2-D
perspective plot. The CI signal displays the average structure of the
correlated emitting sources at kinetic freeze-out. We anticipate kinetic
freeze-out will occur at or near the outer surface of the fireball. The CI
characteristics vs. centrality are shown in Fig. 10 (signal amplitude and 
integral), Fig. 11 ($\Delta \eta$ width), and Fig. 12 ($\Delta \phi$ width).
 
The CI large $\Delta \eta$ elongation, signal amplitude, and signal integral 
all have their maxima in the most central bin (0-5\%) and decrease with 
decreasing centrality. In the most peripheral bins jet-like angular symmetry is
restored and HIJING \cite{Wang} can be expected to fit. However the central bins
strongly reject a HIJING fit due to the $\Delta \eta$ elongation which does not
have a jet-like angular symmetry which HIJING requires. Thus only the last few 
peripheral bins are consistent with HIJING. HIJING can not account for the 
strength of the $\Delta \eta$ and $\Delta \phi$ correlations in the most 
central collisions see Ref. \cite{PBM}.
 
The CI $\Delta \phi$ width remains approximately constant with centrality
($\sim30^\circ$). This implies the source we are viewing covers a fraction of
the $\phi$ angular range at all centralities in our observed $p_t$ range.
This was calculated in Ref. \cite{themodel} and shown to be so. In our $p_t$ 
range the limited angular spread of particles in $\Delta \phi$ could be due 
to phase space focusing by flow as considered by Ref. \cite{HBT} as
an explanation of the HBT results. A blast wave fit including all pions gives 
source sizes which are consistent with the HBT results for Au + Au central 
collisions at $\sqrt{s_{NN}}$ = 200 GeV \cite{HBT}. 
 
In the blast wave fit used in the PBM the surface of the blast wave is
moving with maximum velocity (3c/4) at kinetic freeze-out when the particles
are emitted. This phase space focusing of source size in the blast wave surface
with the addition of bubble substructure on or near the surface of the blast 
wave is used in the PBM \cite{PBM}. This model assumes a bubble substructure of 
about a dozen similar adjoining spherical bubbles in an 8 fm radius ring at 
mid-rapidity, perpendicular to and centered around the beam direction, on the 
surface of the fireball at kinetic freeze-out. The correlation functions used 
in our analysis and those employed for HBT both add all the bubbles in the ring
on top of each other resulting in one average bubble with a source size of 
$\sim$2fm radius. If we were just dealing with flow focusing we would not 
observe the signal correlations we find in this analysis. One needs 
$\Delta \phi$ regions that produce more particles (bubble) and other 
$\Delta \phi$ regions which produce fewer particles (background) in 
order to generate the observed signal correlations. In Ref. \cite{PBM} a 
detailed QCD inspired model was developed which assumed the bubbles were 
composed of gluonic hot spots which appeared in the final state of the fireball
evolution. The PBM reasonably quantitatively fit the CI and the CD of 
the prior central trigger analysis of the approximately 0-10\% 
centrality Au + Au at $\sqrt{s_{NN}}$ = 200 GeV for the intermediate 
transverse momenta 0.8$ < p_t <$ 2.0 GeV/c.

\subsection{Model fits for the CI}
 
The characteristics of the present analysis discussed above suggest that the 
PBM could fit the two most central CI (total correlation) bins and HIJING could
fit the most peripheral bins. It appears that the in between bins would require
a mixture of the PBM, HIJING, and elliptic flow for a reasonable fit. Future 
theoretical work is expected to address fitting the in between bins.

Let us first address fitting the two CI (total correlation) most central bins 
(0-5\% and 5-10\%) with PBM and the two most peripheral bins (60-70\% and 
70-80\%) with HIJING. For the two CI most central bins we use the Monte Carlo 
generated PBM events that were used in comparing with the central trigger 
data \cite{centralproduction}. The entire CI correlation (signal + background) 
is used for comparing the analysis results with the model. This eliminates 
any model dependence on the separation of signal from background. In 
Ref. \cite{PBM} the centrality range was 0-10\%, while here we must separate 
the Monte Carlo sample into two pieces (0-5\% and 5-10\%).

The correlation derived when using all particles to form pairs regardless of 
their charge sign is equal to CI/2. Since our correlations are normalized to a 
mean of 1, plotting (CI/2 - 1) provides a more easily understood theoretical 
comparison in Fig. 20.

We need to use multiplicity scaling in order to compare the 
different centralities. This comparison was straightforward for exhibiting the
2-D fits of the CI signal vs. centrality (see Fig. 8) but becomes much harder 
when comparing the total scaled CI with models. For each centrality 
the CI/2 varies about 1.0. Since it is very hard to compare 2-D correlations 
we project into a 1-D presentation by dividing 1-D $\Delta \phi$ projections 
into five $\Delta \eta$ ranges $0.0 < \Delta \eta < 0.3$, $0.3 < \Delta \eta 
< 0.6$, $0.6 < \Delta \eta < 0.9$, $0.9 < \Delta \eta < 1.2$, and 
$1.2 < \Delta \eta < 1.5$ which cover our entire $\Delta \eta$ range.

Fig. 20 compares the increasing $\Delta \eta$ sequence to the PBM
fits (lines) in the two most central bins and to the HIJING fits
(dashed lines) in the two most peripheral bins.  The horizontal axis is 
$\Delta \phi$ in degrees. 

The comparison to HIJING jets is not shown for the two most central bins 
(0-5\%) and (5-10\%) because HIJING has a large disagreement with the data. 
This is due to the large observed $\Delta \eta$ elongation (by a factor of
$\sim$3 see Section IV A) in the CI data which is not contained in HIJING
jets. This contradiction of HIJING in the most central bins is shown in Fig. 17
of Ref. \cite{PBM} where the centrality range is 0-10\%.

The away side of the scaled correlation ($\Delta \phi$ = $90^\circ$ - 
$180^\circ$) is mainly $v_2$ plus some contribution from 
$\rm\cos$($\Delta \phi$) which is close to the entire background. Thus the near
side CI signals ($\Delta \phi$ = $0^\circ$ -$90^\circ$) which are the signals 
we have extracted and analyzed in Section IV A should rise above the value of
the away side. This clearly occurs in each $\Delta \eta$ bin shown in Fig. 20.  

\begin{figure*}[ht] \centerline{\includegraphics[width=0.800\textwidth]
{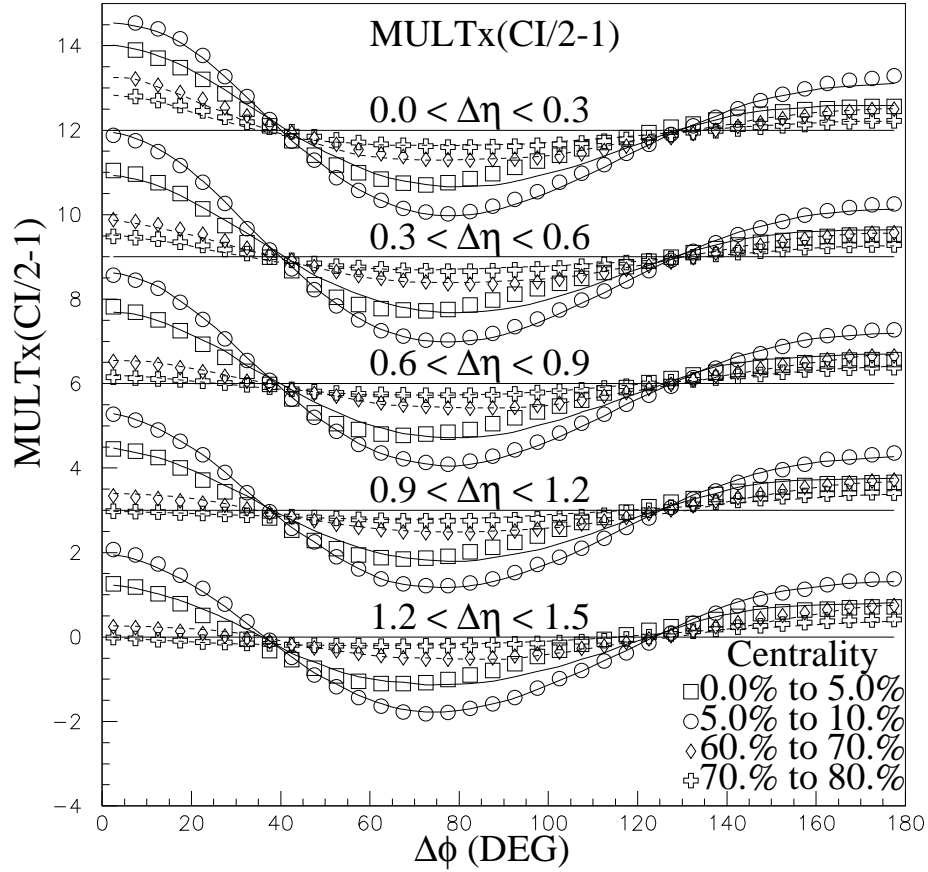}} \caption[]{For the $\Delta \eta$ ranges
$0.0 < \Delta \eta < 0.3$, $0.3 < \Delta \eta < 0.6$,
$0.6 < \Delta \eta < 0.9$, $0.9 < \Delta \eta < 1.2$, and
$1.2 < \Delta \eta < 1.5$ we compare the multiplicity x (CI/2-1) as a function
of  $\Delta \phi$ (degrees) with the parton bubble model (PBM) fits (lines)
for 0-5\% and 5-10\% centrality bins. Also a comparison with the HIJING fits
(dashed lines) for 60-70\% and 70-80\% centrality bins is shown. For the 5-10\%
the PBM fit appears to have a larger signal, but this difference is due to
elliptic flow. The data correlations and PBM fits are shifted up by 12 units
for $0.0 < \Delta \eta < 0.3$, 9 units for $0.3 < \Delta \eta < 0.6$,
6 units for $0.6 < \Delta \eta < 0.9$, 3 units for $0.9 < \Delta \eta < 1.2$,
and no shift for $1.2 < \Delta \eta < 1.5$. In comparison with PBM in this
figure we utilize (CI/2 - 1) where the CI correlation = US + LS. The comparison
to HIJING jets is not shown for the two most central bins (0-5\%) and (5-10\%) 
because HIJING has a large contradiction of the data. This is due to the 
large observed $\Delta \eta$ elongation (by a factor of $\sim$3 see Section IV 
A) in the CI data which is not contained in HIJING jets. This contradiction of 
HIJING in the most central bins is shown in Fig. 17 of Ref. \cite{PBM} where the 
centrality range is 0-10\%.}
\label{figure20}
\end{figure*}
                                                                           
\begin{figure*}[ht] \centerline{\includegraphics[width=0.800\textwidth]
{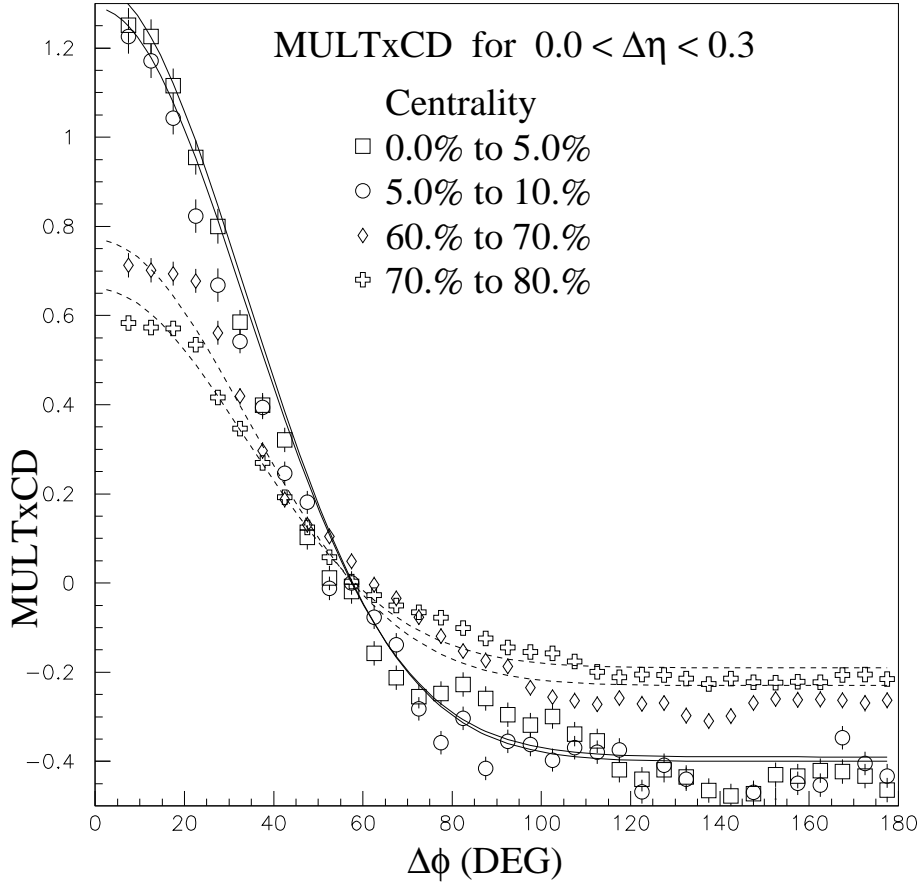}} \caption[]{For the $\Delta \eta$ range
$0.0 < \Delta \eta < 0.3$ a comparison of Multiplicity x CD analysis results as
a function of  $\Delta \phi$ (degrees) with the parton bubble model fits
(lines) for 0-5\% and 5-10\% centrality bins. Also a comparison with the HIJING
fits (dashed lines) for 60-70\% and 70-80\% centrality bins is shown.}
\label{figure21}
\end{figure*}
                                                            
\begin{figure*}[ht] \centerline{\includegraphics[width=0.800\textwidth]
{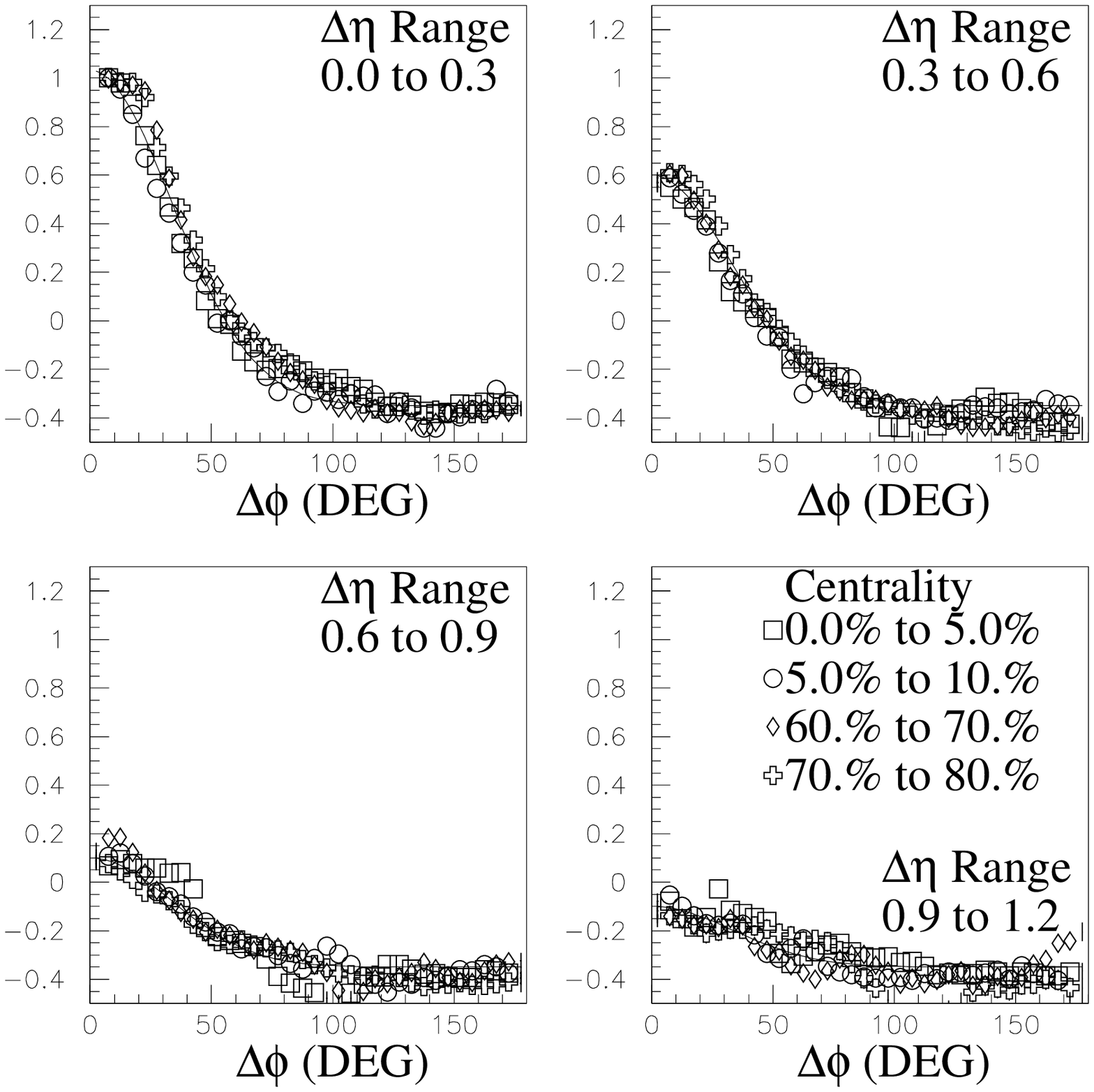}} \caption[]{The product of the Multiplicity
and the CD correlation vs. $\Delta \phi$ for 4 $\Delta \eta$ bins. Four
$\Delta \eta$ ranges with four centralities each are shown 0-5\%, 5-10\%,
60-70\%, and 70-80\%. Each of the centralities were scaled so that the
$5-10^\circ$ $\Delta \phi$ bin for the $\Delta \eta$ range 0.0 to 0.3 is
normalized to 1. In each of the four $\Delta \eta$ ranges we see that all 4
centrality bins cluster around the Pythia jets lines shown. Thus we see that
the CD shape is approximately independent of centrality and the CD is
approximately consistent with Pythia jets \cite{pythia} (dashed lines) at each
centrality for all $\Delta \eta$ ranges shown. This is true for all nine CD
centrality bins (see Section IV B last paragraph), and thus is consistent with
surface emission from the fireball at kinetic freeze-out.}
\label{figure22}
\end{figure*}
                                                       
From Fig. 20 it is clear that the PBM fits for the 0-5\% and 5-10\%  
bins agree with the CI analysis results within an average of a few percent
of the total correlation for all $\Delta \eta$ ranges. HIJING fits agree with 
the analysis results for the CI 60-70\% and 70-80\% bins within an 
average of a few \% for all $\Delta \eta$ ranges. Therefore the entire  2-D 
analyzed CI correlations, which are the sum of the five $\Delta \eta$ ranges, 
agree with the PBM fits for the 0-5\% and 5-10\% bins within
an average of a few percent of the CI total correlations. The 2-D analyzed CI 
correlations agree with HIJING fits for the 60-70\% and 70-80\% 
bins within an average of a few percent of the CI total correlations. As we
shall see in the following subsection the CD of the PBM agrees with the CD 
analysis results with similar precision (within a few percent of the 
correlation). 

\subsection{Model fits for the CD}

We again use the CD (total correlation) for comparison with
theoretical models. Fig. 21 shows a comparison of Multiplicity x CD analysis 
results as a function of $\Delta \phi$ (degrees) with the PBM  fits 
(lines) for 0-5\% and 5-10\% centrality bins for the $\Delta \eta$ range
$0.0 < \Delta \eta < 0.3$. Also shown is a comparison with the HIJING fits 
(dashed lines) for the 60-70\% and 70-80\% centrality bins. From 
the above comparison we observe that the PBM fits for the 0-5\% and 5-10\% 
centrality bins agree within a few \% with the CD analysis results. The same is
also true for the HIJING fits 60-70\% and 70-80\%. 

In Fig. 22 we plot the Multiplicity x CD as a function of $\Delta \phi$ 
(degrees) for 4 $\Delta \eta$ bins covering the range 
$0.0 < \Delta \eta < 1.2$. Each of the above four centralities shown was 
scaled so that the $5-10^\circ$ $\Delta \phi$ bin for $0.0 < \Delta \eta < 0.3$
is normalized to 1. We did this to show that the CD shape is 
approximately independent of centrality. This is consistent with Fig. 18-19
which show the Gaussian rms width in $\Delta \phi$ and $\Delta \eta$ for the
CD signal. One can see in Fig. 22 that the analysis points in each $\Delta \eta$
range cluster around the four lines which correspond to each of the four 
$\Delta \eta$ ranges generated by Pythia jets \cite{pythia}. 

Pythia jets are used as the jets in HIJING and Pythia fragmentation is used in 
the PBM; thus the CD shape is totally given by Pythia and is approximately 
independent of centrality (see Fig. 18-19 ). If appreciable further interaction 
with the fireball interior occurred these initial CD correlations would 
be changed. Therefore we can conclude that the emitted pairs 
have little further interactions after hadronization. Thus surface or 
near surface hadronization and emission from the fireball both occur in the 
central region and all other centralities where there is considerable particle 
density. In the most peripheral bins the particle density is low enough to 
allow undisturbed fragmentation and thus no change in the CD correlation.
It should be noted that the CD of the PBM in the 0-5\% and 5-10\% centrality
bins is also reasonably consistent (within a few percent of the correlation)
with Pythia jets. Thus the CI and the CD of the PBM in the 0-5\% and 5-10\% 
centrality bins are in reasonable quantitative agreement with the data 
analysis.

A theoretical pQCD calculation \cite{kajantie} concluded that minijets 
formed with initial parton transverse momenta of around 3 GeV/c (also 
applicable down to 2 GeV/c) would become thermalized in a 200 GeV/c U + U 
collision at RHIC and likely not escape from the system. However the incomplete
treatment of the fireball medium in this model \cite{kajantie} could raise some 
doubt about the calculation. The authors freely admit that they cannot use pQCD
for softer minijet calculations. Nevertheless their pQCD calculations show that
the minijets formed by initial hard scatterings should be thermalized using 
reasonable cross sections. Thus the initial correlations from minijets are not 
expected to survive to the final state. 

This minijet model \cite{kajantie} has been modified and utilized as a 
qualitative picture to explain STAR correlation data by assuming that the 
observed correlations are surviving correlations generated by initial hard 
scatterings that interact with the medium of the fireball interior. The 
correlations have only been modified (accounting for the change in the CI
$\Delta \eta$ width see Fig. 11) and not destroyed \cite{aya}. This minijet model
qualitative picture is totally different from the results of the pQCD
treatment just discussed which concluded the minijets are thermalized and 
do not produce the observed correlations.

The fact that in the $p_t$ range 0.8 GeV/c$ < p_t <$ 4.0 GeV/c of this
analysis the CD correlation shape is consistent with Pythia jets fragmentation 
becomes a serious challenge for this minijet model \cite{aya}. Since the 
behavior of the CD in our $p_t$ range is consistent with both hadronization 
and kinetic freeze-out occurring at or near the fireball surface it is clearly 
implied that initial state hard scattering as assumed in this minijet 
model \cite{aya} does not survive to the final state at freeze-out. Therefore we
conclude this and likely any  minijet model assuming survival of initial 
hard scattering to the final state correlations fails in our $p_t$ range.

\section{Summary and Conclusions}

We performed an experimental investigation of particle-pair correlations in 
$\Delta \phi$ and $\Delta \eta$ using the main Time Projection Chamber of the 
STAR detector at RHIC. We investigated Au + Au collisions at 
$\sqrt{s_{NN}}$ = 200 GeV as a function of centrality using 13.5 million events
from a minimum bias trigger to select the nine 0-80\% centrality bins. In order
to allow resolution of the $\sim$2fm radii substructures we cut out the lower 
momentum particles which come from sources that have radii up to 6 fm in size. 
These lower momentum particles would degrade our resolution of the $\sim$2fm 
sources and add particles which came from interacting in the interior of the 
fireball. Treatment of such interacting particles is complex. Thus we selected 
tracks having transverse momenta $0.8 < p_t < 4.0$ GeV/c, and $|\eta| < 1.0$. 
We performed cuts to remove the small angle bins necessary to reduce to a 
negligible level the effects of track merging, Coulomb, and HBT. 
Using symmetries of the data in $\Delta \eta$ and $\Delta \phi$ we 
were able to fold the four quadrants into one. The entire data set, 
unlike-sign charge pairs (US) and like-sign charge pairs (LS) for each 
centrality bin, was fit by a reasonably interpretable set of parameters. 
These consisted of 13 parameters for the  US and 15 for the LS. These 
parameters are small in number compared to the total number of 
$\Delta \phi$ $\Delta \eta$ bins (534 - 536) used in each centrality bin. 

One should note that the LS signal consisted of two Gaussians and required 
only 6 signal parameters. The US signal consisted of one approximate 
Gaussian which had an additional term in the exponent which depended on the 
fourth power of $\Delta \eta$. Therefore four parameters were used for the US 
signal. 

In order to fit the background in this precision experiment and represent 
known and expected physics effects, 9 parameters were required for
both US and LS. The background terms are similar but not identical in US and LS
as expected. However in order to get the best precision and minimum distortion 
in the fits they were kept separate. The physical arguments for the 
necessity of these parameters were given in Section III.

All the fits to the US and LS for all centrality bins had significance
consistent with $3\sigma$ or less. The parameterization used in this 
analysis was very similar to that successfully used in the prior 
central trigger analysis \cite{centralproduction}. The major differences were 
the removal of the readout sector gap terms and an additional flow term
$\rm\cos$($4\Delta \phi$). Appendix A contains the parameters and their 
variation with centrality.

The Charge Independent (CI $\equiv$ US + LS) and the Charge Dependent 
(CD $\equiv$ US - LS) signals were formed for each of the nine centrality 
bins. The CI signal displays the average structure of the correlated 
emitting sources at kinetic freeze-out. Thus the total CI correlation is an 
unbiased quantitative measure of the particle pair correlation observed in 
the TPC. The CI signal characteristics are shown and discussed in 
Section IV A. The most central CI bins have the largest signals and a 
large $\Delta \eta$ elongation. Both the signal and elongation 
decrease with deceasing centrality till the most peripheral 
bins where the $\Delta \eta$ and $\Delta \phi$ correlation distributions 
are symmetric. The PBM fits the two most central bins very well within 
an average of a few \% of the total correlation. HIJING fits the two most 
peripheral bins with similar precision (See Section VI B). Future 
theoretical work is expected to address fitting the intermediate 
centrality region of the CI and to investigate the source characteristics 
in these intermediate centrality bins.

Section V discusses systematic errors. From our analysis of the systematic 
errors we conclude that the conclusions drawn have not been significantly 
effected by the systematic errors.

An overall interpretation that is consistent with this analysis is:

1) The CD correlation would represent the intial correlation of opposite sign 
charge pairs emitted from the same space-time region if the emission occurred
from, or very near to, the surface of the fireball at kinetic freeze-out. 
It was shown in Section VI C that the shape of the CD correlation is 
approximately independent of collision centrality and consistent with HIJING 
(or Pythia) jets. Thus, we conclude that both hadronization and particle 
emission are consistent with coming from, or very near to, the surface. At
peripheral centralities, the intermediate $p_t$ range charged particles are
mainly produced by jets which fragment freely. As the centrality increases,
a fireball forms which may be very complex, but the amount of fireball 
material that the intermediate $p_t$ particles pass through is minimal. The 
justification for this is the approximate invariance of the shape of the CD 
correlation with respect to centrality. Furthermore, the Pythia jet shape
for the intermediate $p_t$ charged particles is consistent with the measured 
CD correlation for all centralities.

2) In the most central bins, the simultaneous peaking of the 2-D CI signal 
amplitude, signal integral, and the large $\Delta \eta$ elongation which occurs
and the CD are all well fit by the parton bubble model. The PBM fits the two 
most central bins well within an average of a few percent. The model was 
developed to be consistent with the HBT observed source size. There is
a minijet model based on pQCD calculations for 200 GeV U + U collisions at
RHIC which considered minijets formed with partons of transverse momenta of 
around 3 GeV/c \cite{kajantie}. The authors concluded that the minijets would 
become thermalized and likely not escape from the system. In Section VI C we 
discussed the general characteristics of minijet correlations (details and 
limitations) originating from hard scatterings which may be observed. The
precision analysis of the data as a function of centrality presented in
this paper could stimulate other new physics models as possible explanations
of the observed correlations.

\section{Acknowledgment} 

We thank the RHIC Operations Group and RCF at BNL, and the
NERSC Center at LBNL and the resources provided by the
Open Science Grid consortium for their support. This work 
was supported in part by the Offices of NP and HEP within 
the U.S. DOE Office of Science, the U.S. NSF, the Sloan 
Foundation, the DFG Excellence Cluster EXC153 of Germany, 
CNRS/IN2P3, RA, RPL, and EMN of France, STFC and EPSRC
of the United Kingdom, FAPESP of Brazil, the Russian 
Ministry of Sci. and Tech., the NNSFC, CAS, MoST, and MoE 
of China, IRP and GA of the Czech Republic, FOM of the 
Netherlands, DAE, DST, and CSIR of the Government of India, 
Swiss NSF, the Polish State Committee for Scientific Research,
Slovak Research and Development Agency, and the Korea Sci. 
\& Eng. Foundation.

\appendix

\section{Parameters}

The parameterization was similar to that employed in 
Ref. \cite{centralproduction}. 

There are two basic pairs of particles: unlike-sign charge pairs (US) and 
like-sign charge pairs (LS). Both were parameterized for each centrality bin 
independently. The folded after cuts US data as a function of centrality 
shown in Fig. 1 was parameterized using the US parameterization. The folded 
after cuts LS data as a function of centrality shown in Fig. 2 was 
parameterized using the LS parameterization. Thus there were nine independent 
US and LS sets of parameters.

The significant difference from Ref. \cite{centralproduction} run 2 parameters 
were:

1) The sector gap terms were not needed because there was event by event
space charge distortion correction for this run (run 4). In 
Ref. \cite{centralproduction} we had attributed the necessity of the sector gap 
corrections to space charge. This explains why they were not necessary in 
run 4.

2) We found it necessary to include a $\rm\cos$($4\Delta \phi$) flow term in 
addition to the previously employed $2 v_2^2\rm\cos$($2\Delta \phi$) term to 
obtain good fits.

Parameters fitting requires a mixture of knowledge and experience. One must 
keep in mind several important points.

a) One must use one's knowledge and experience to pick realistic parameters 
which will efficiently and meaningfully describe the data well enough so that 
the statistical significance of the final fits are good enough to be 
credible. 

b) Experienced data analysts usually require that their fits be consistent with
at least a $3\sigma$ fit compared to an ideal fit in which $\chi^2$ =
degrees of freedom (DOF). 

\subsection{Parameters of the background terms for US and LS vs. Centrality}

The first background term is an overall normalization term of the correlation
to a mean of 1 shown in Fig. 23.

Four terms Fig. 24-27 are needed for momentum and charge conservation in order
to obtain acceptable fits.

Two terms Fig. 28-29 represent elliptic flow effects and are background 
parameters as far as this analysis is concerned. No reaction plane is assumed
in this analysis.

Fig. 30 is a background term probably due to remaining long range correlation. 
Long range correlations should be $\Delta \phi$ independent like this term and 
important in the soft particle $p_t$ range which we do not explore in this 
analysis. 

Fig. 31 is a $\phi$ independent small effect which we attribute to losses in
the larger $\eta$ tracking in the TPC and perhaps part of a long range
correlation not relevant to this analysis. We utilized mixed-event-pairs with 
a similar $z$-vertex to take into account these losses. Imperfections in this 
procedure leave a small Gaussian bump near larger $\Delta \eta$. We found that
choosing a Gaussian with a fixed center of 1.25 and a fixed width of 1.57 for 
all centrality bins was adequate. 

\subsection{Signal Parameters for the US vs. Centrality}

Fig. 32 is the US signal amplitude for the approximate Gaussian fit.

Fig. 33 is the US signal $\Delta \eta$ width for the approximate Gaussian fit.

Fig. 34 is the US signal additional term $(\Delta \eta)^4$ in the Gaussian 
exponent which makes it an approximate Gaussian fit.

Fig. 35 is the US signal $\Delta \phi$ width for the approximate Gaussian fit.

\subsection{Signal Parameters for the LS vs. Centrality}

Fig. 36 is the LS signal amplitude for the large Gaussian in the fit.

Fig. 37 is the LS signal $\Delta \eta$ width for the large Gaussian in the fit.

Fig. 38 is the LS signal $\Delta \phi$ width for the large Gaussian in the fit.

Fig. 39 is the LS signal amplitude for the dip Gaussian in the fit.

Fig. 40 is the LS signal $\Delta \eta$ width for the dip Gaussian in the fit.

Fig. 41 is the LS signal $\Delta \phi$ width for the dip Gaussian in the fit.

\begin{figure*}[ht] \centerline{\includegraphics[width=0.800\textwidth]
{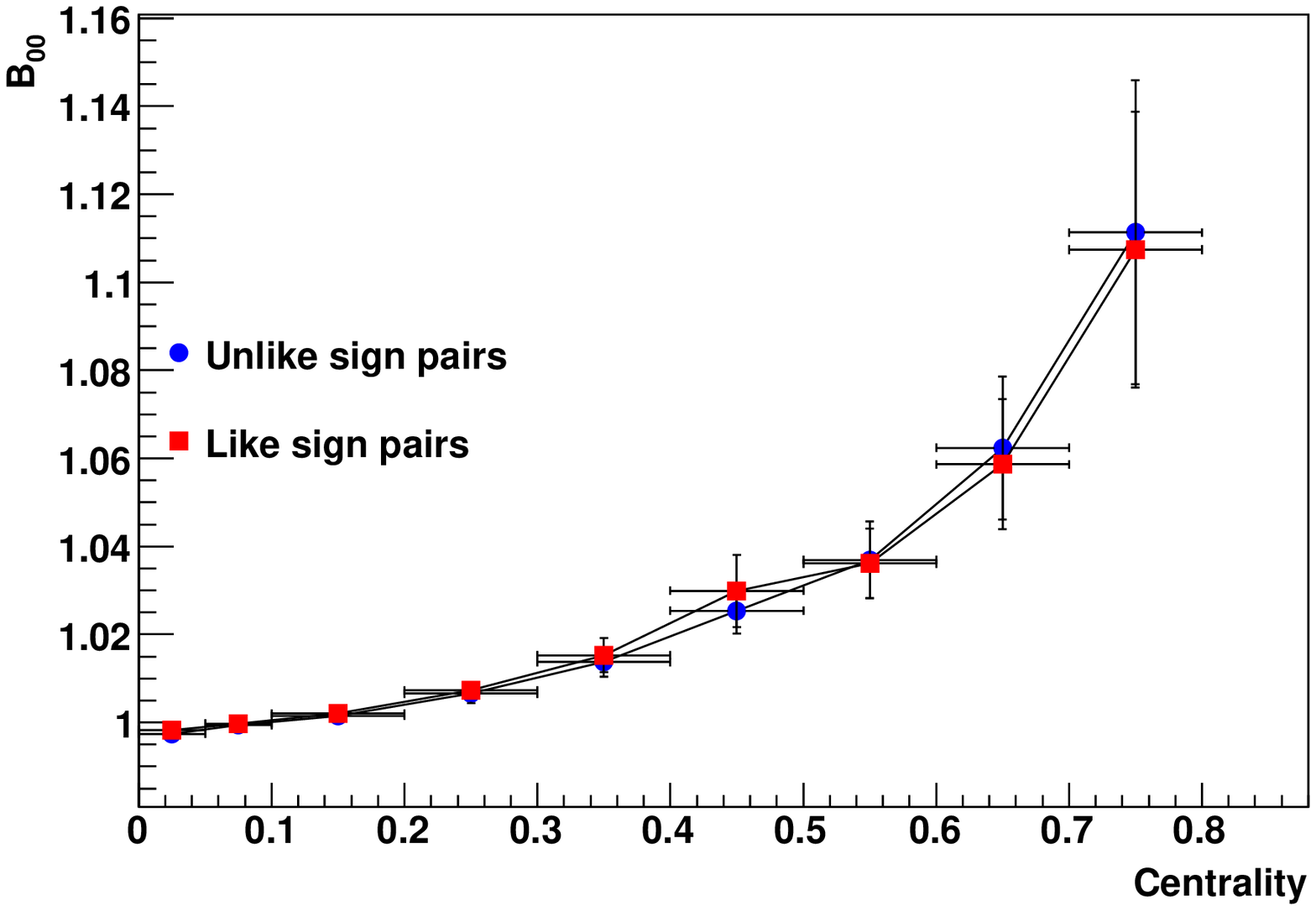}} \caption[]{``(Color online)'' The normalization constant
compared for US and LS.}
\label{figure23}
\end{figure*}
                                                                           
\begin{figure*}[ht] \centerline{\includegraphics[width=0.800\textwidth]
{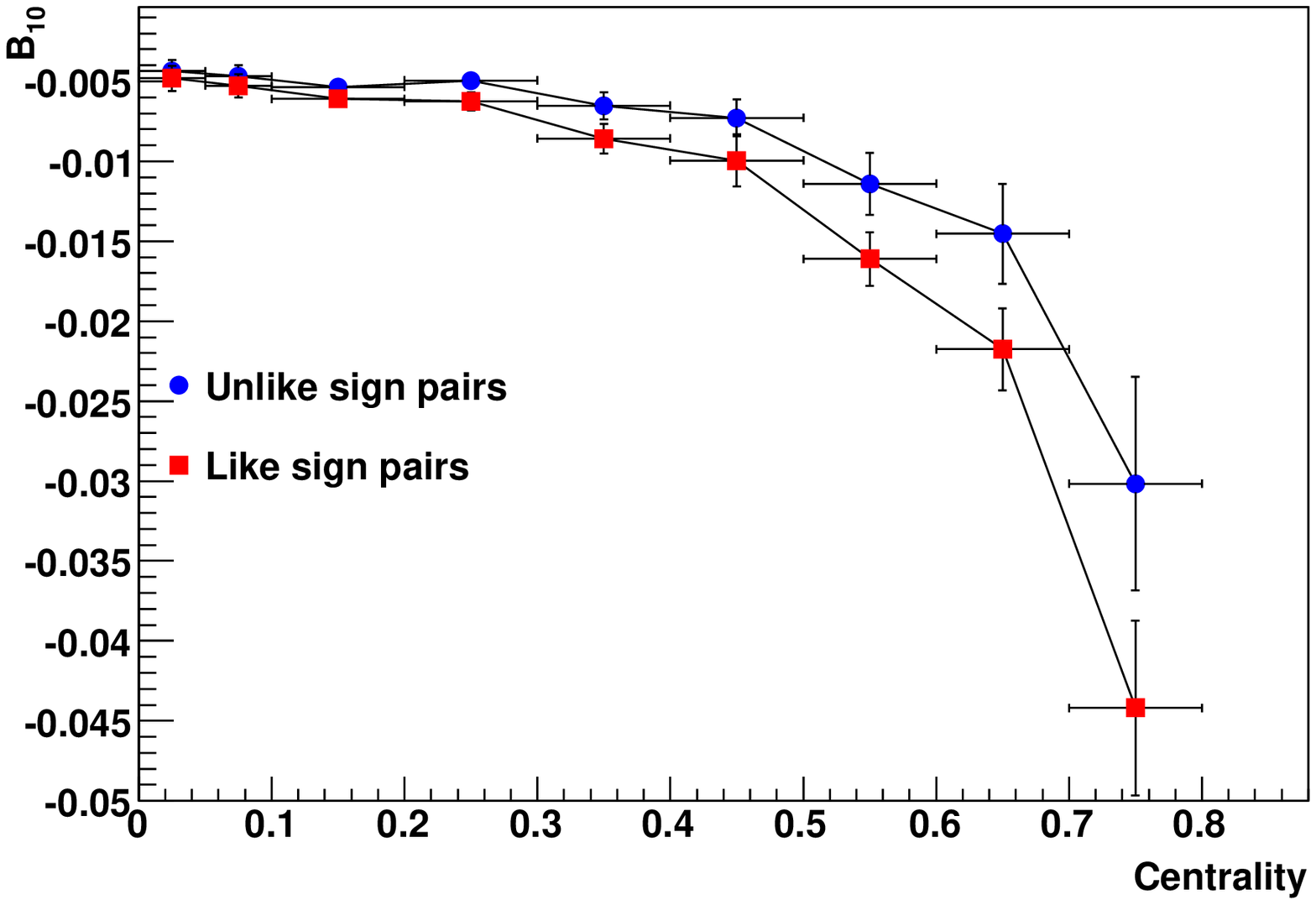}} \caption[]{``(Color online)'' The $\cos(\Delta \phi)$ term
compared for US and LS.}
\label{figure24}
\end{figure*}

\begin{figure*}[ht] \centerline{\includegraphics[width=0.800\textwidth]
{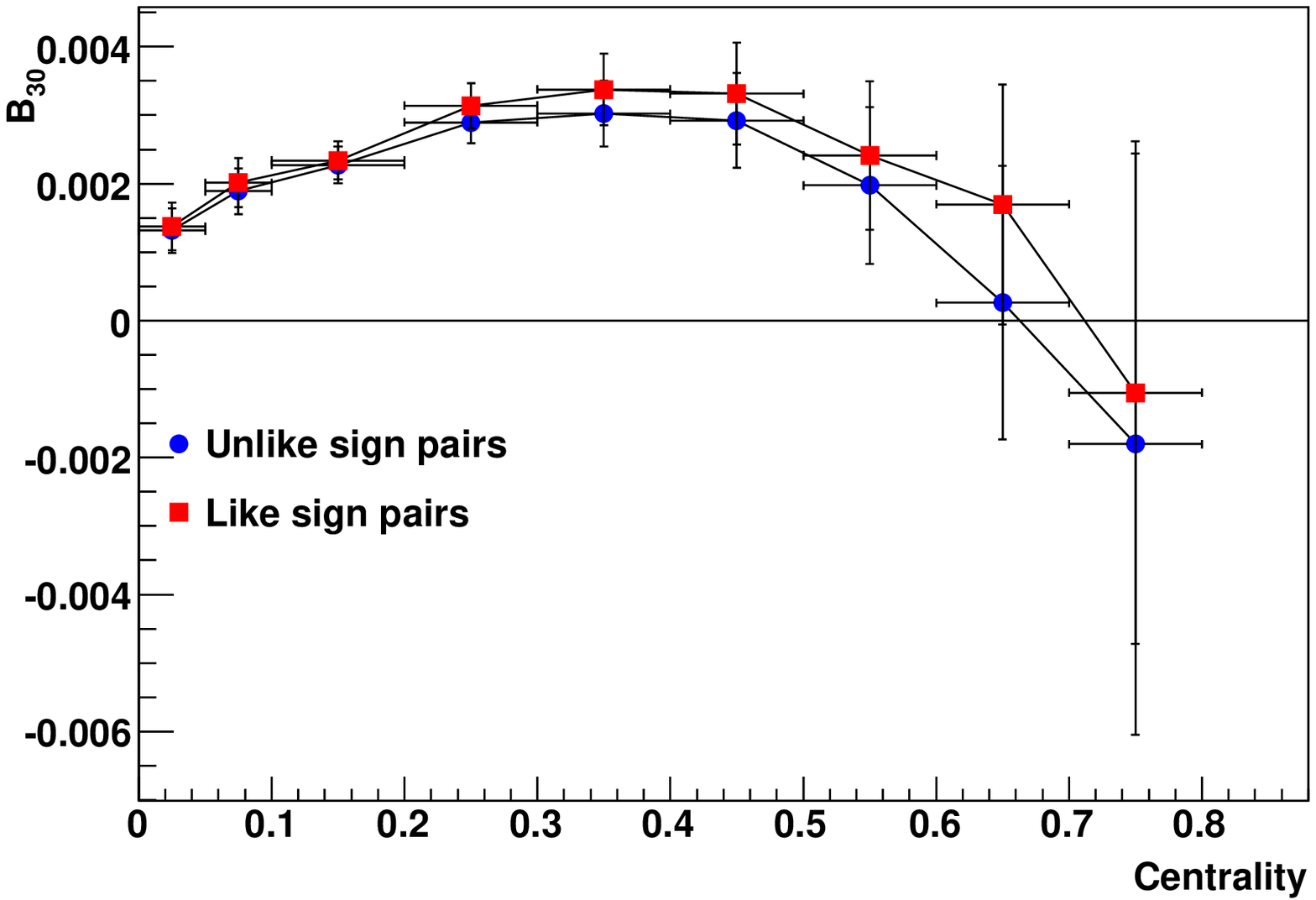}} \caption[]{``(Color online)'' The $\cos(3\Delta \phi)$ term
compared for US and LS.}
\label{figure25}
\end{figure*}
                                                                          
\begin{figure*}[ht] \centerline{\includegraphics[width=0.800\textwidth]
{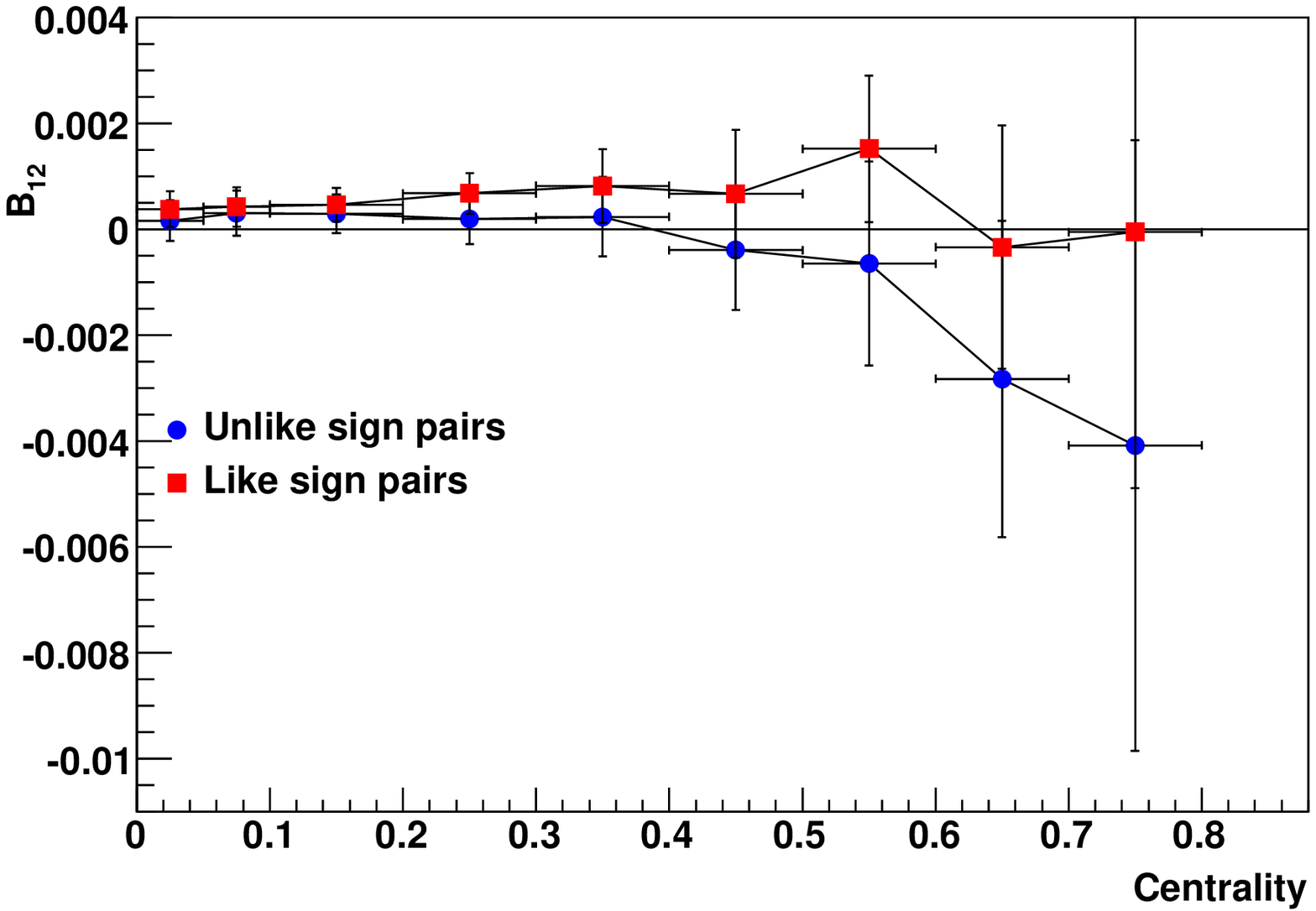}} \caption[]{``(Color online)'' The
$(\Delta \eta)^2\cos(\Delta \phi)$ term compared for US and LS.}
\label{figure26}
\end{figure*}
                                                                            
\begin{figure*}[ht] \centerline{\includegraphics[width=0.800\textwidth]
{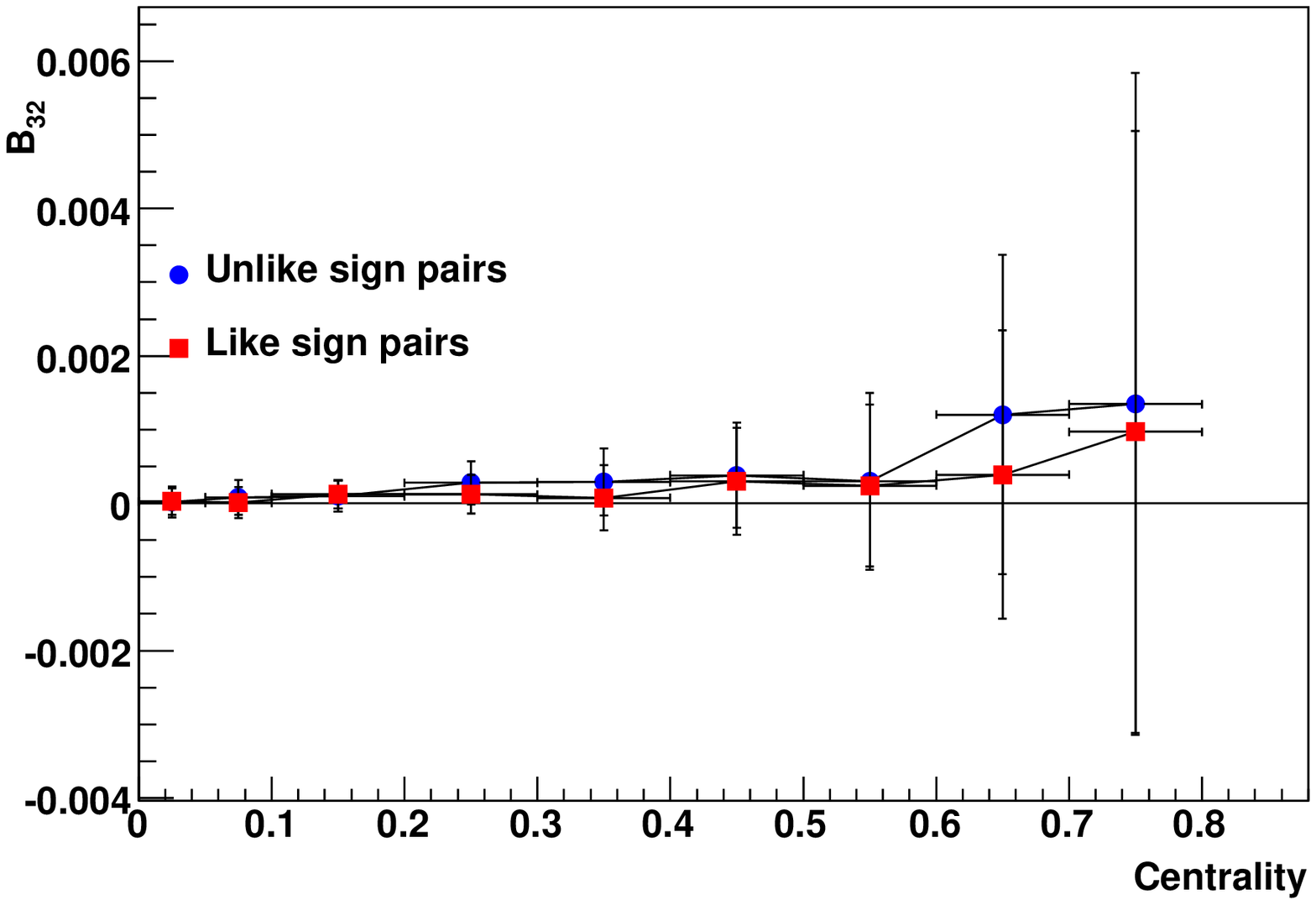}} \caption[]{``(Color online)'' The
$(\Delta \eta)^2\cos(3\Delta \phi)$ term compared for US and LS.}
\label{figure27}
\end{figure*}

\begin{figure*}[ht] \centerline{\includegraphics[width=0.800\textwidth]
{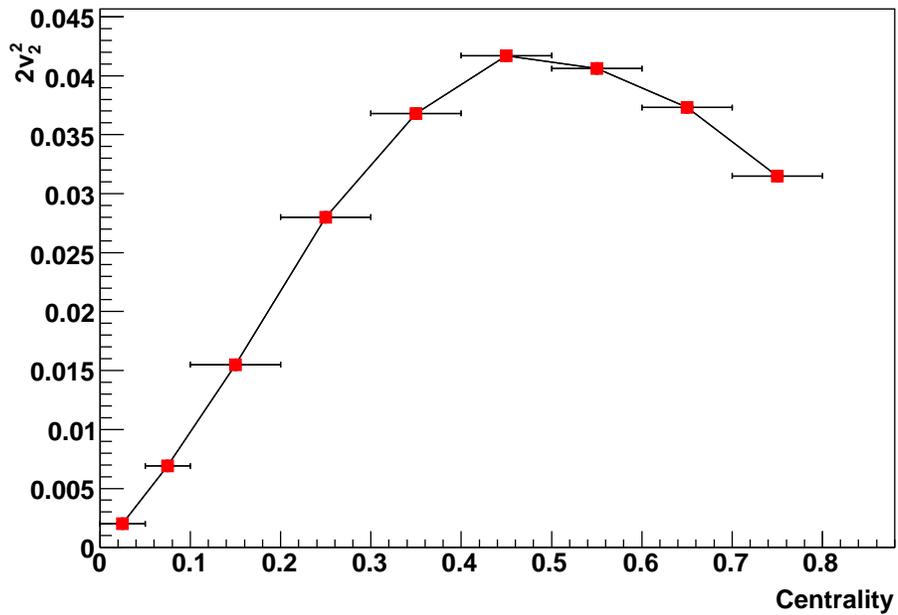}} \caption[]{``(Color online)'' The $v_2$ term plotted is actually
$2 v_2^2$ and is constrained to be the same for both US and LS.}
\label{figure28}
\end{figure*}
                                                                  
\begin{figure*}[ht] \centerline{\includegraphics[width=0.800\textwidth]
{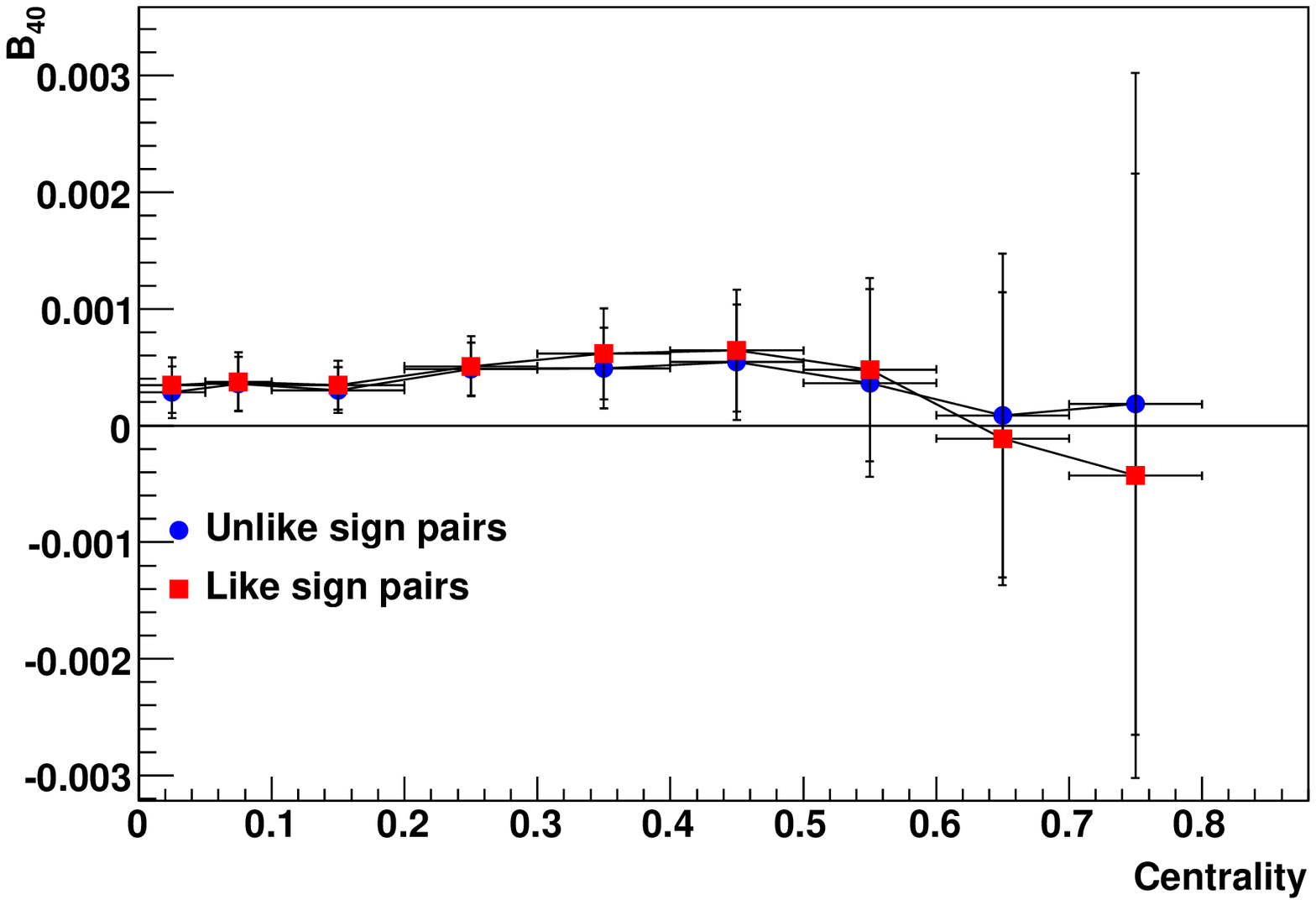}} \caption[]{``(Color online)'' The $\cos(4\Delta \phi)$ term
compared for US and LS.}
\label{figure29}
\end{figure*}
                                  
\begin{figure*}[ht] \centerline{\includegraphics[width=0.800\textwidth]
{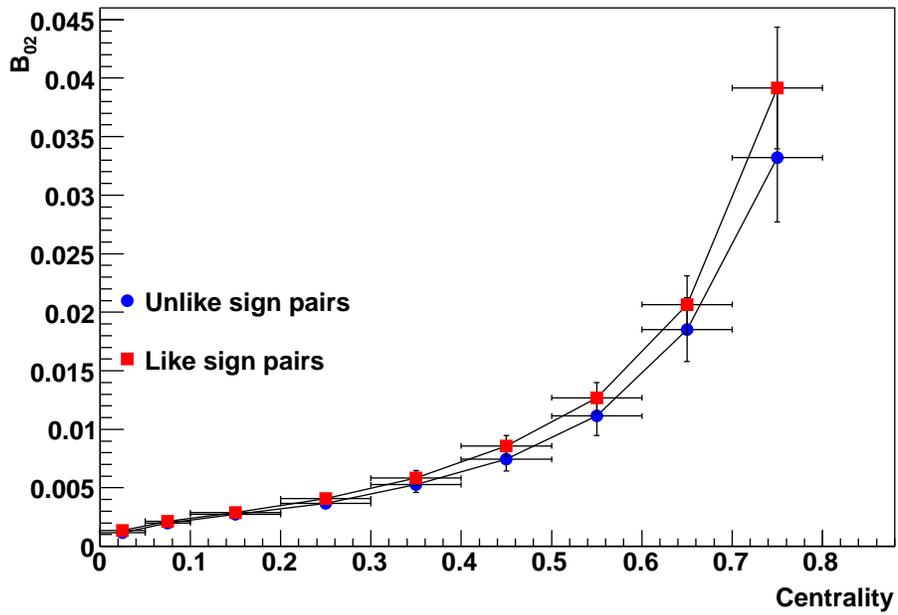}} \caption[]{``(Color online)'' The $(\Delta \eta)^2$ term
compared for US and LS. This term is probably due to remaining long range
correlation.}
\label{figure30}
\end{figure*}
                                                                             
\begin{figure*}[ht] \centerline{\includegraphics[width=0.800\textwidth]
{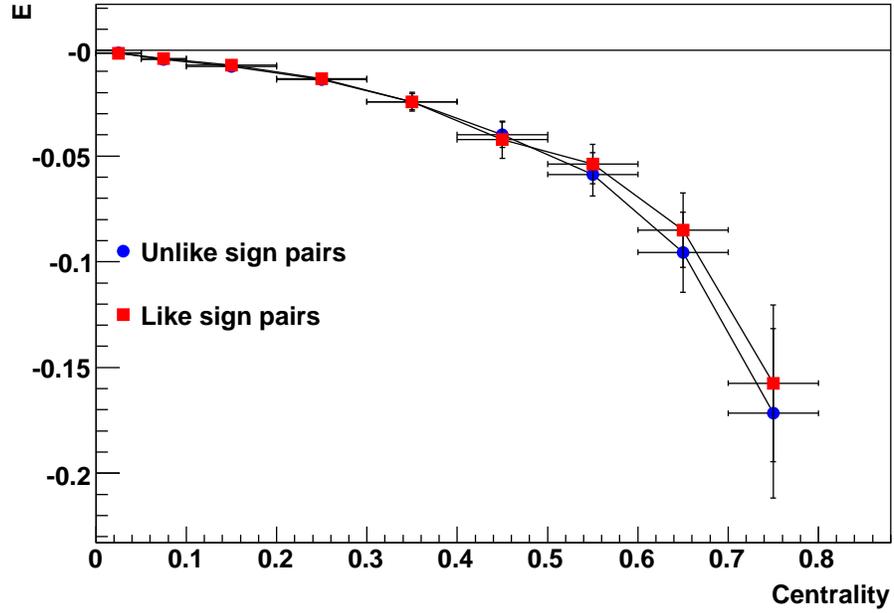}} \caption[]{``(Color online)'' A small $\Delta \phi$ 
independent effect is compared for US and LS. This small term is attributed to 
a correction for the large $\eta$ tracking effect in the TPC which remains 
after division by mixed pairs, and perhaps part of a longe range correlation 
not relevant to this analysis}
\label{figure31}
\end{figure*}
                                                                            
\begin{figure*}[ht] \centerline{\includegraphics[width=0.800\textwidth]
{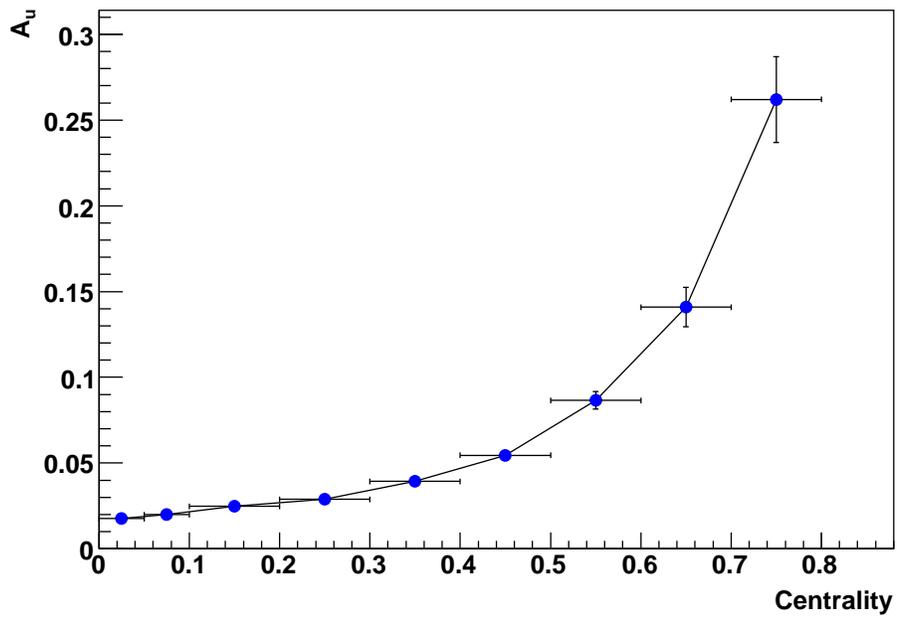}} \caption[]{``(Color online)'' The US signal amplitude for the
approximate Gaussian fit.}
\label{figure32}
\end{figure*}
                                                                   
\begin{figure*}[ht] \centerline{\includegraphics[width=0.800\textwidth]
{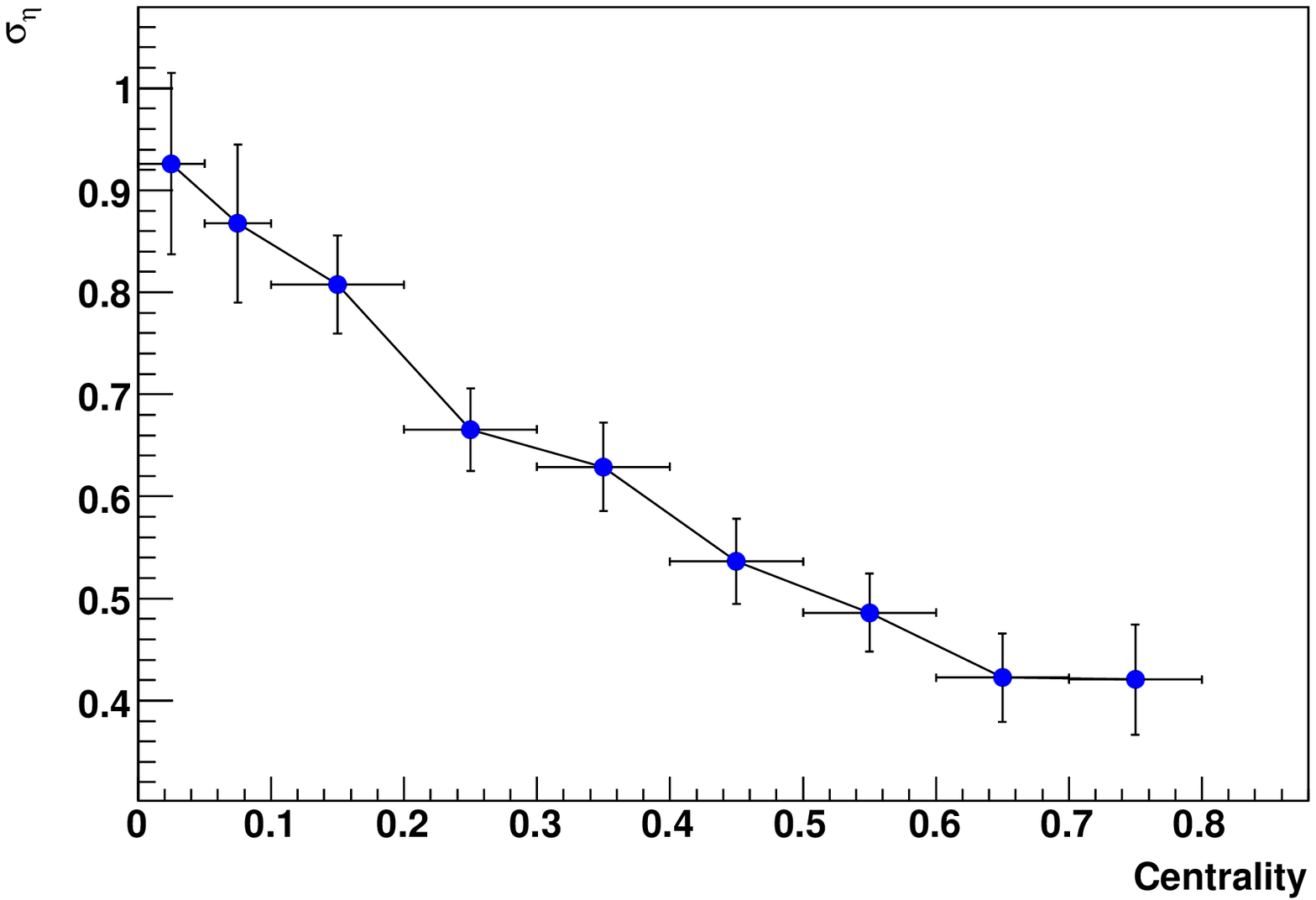}} \caption[]{``(Color online)'' The US signal $\Delta \eta$
width for the approximate Gaussian fit.}
\label{figure33}
\end{figure*}
                                    
\begin{figure*}[ht] \centerline{\includegraphics[width=0.800\textwidth]
{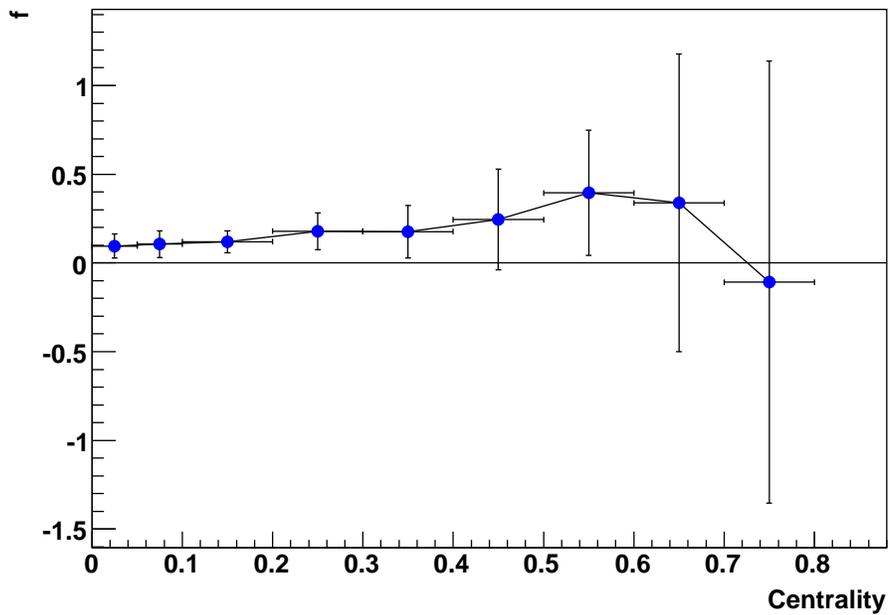}} \caption[]{``(Color online)'' the US signal additional term
$(\Delta \eta)^4$ in the Gaussian exponent which makes it an approximate
Gaussian fit.}
\label{figure34}
\end{figure*}
                                                                             
\begin{figure*}[ht] \centerline{\includegraphics[width=0.800\textwidth]
{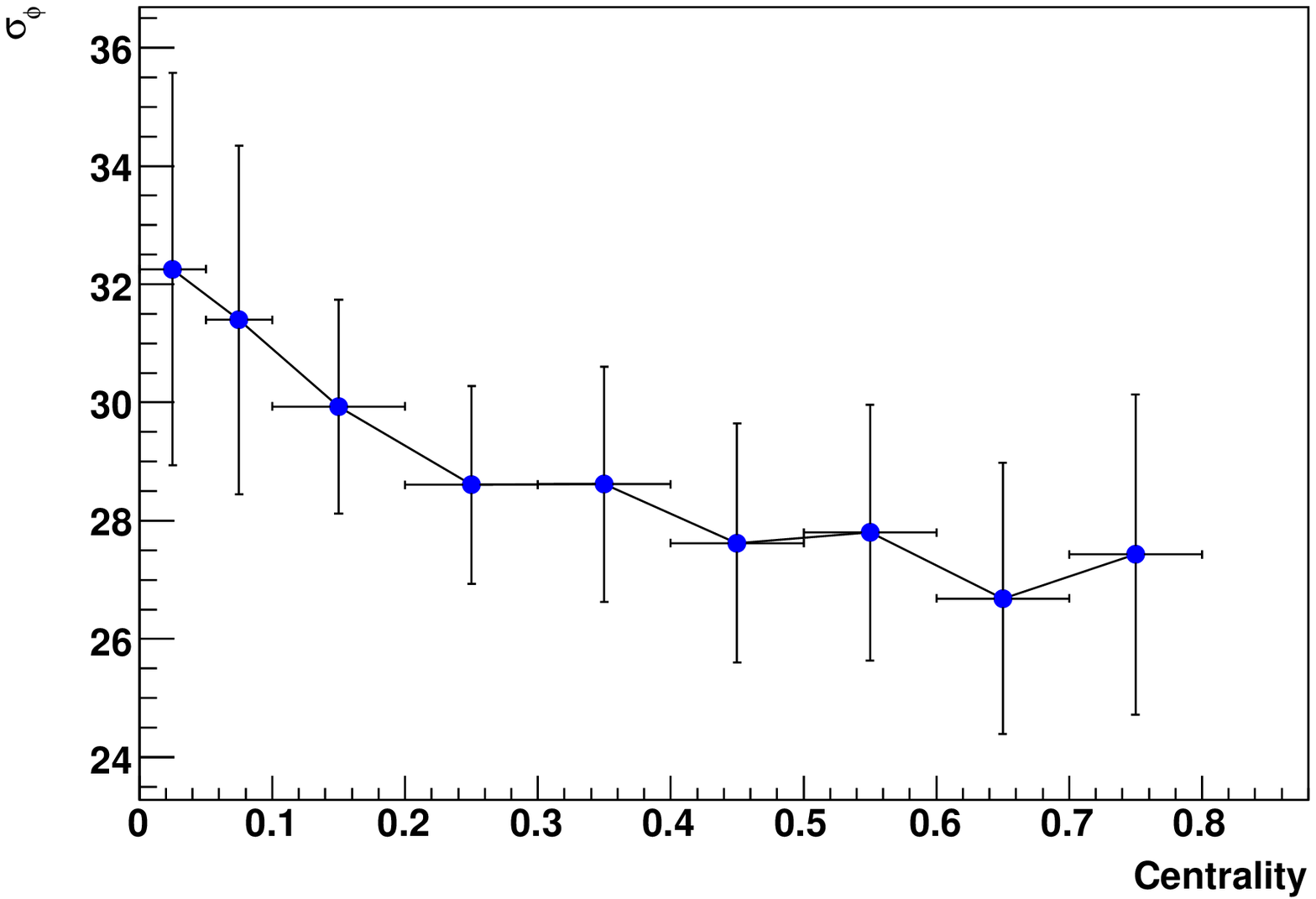}} \caption[]{``(Color online)'' The US signal $\Delta \phi$
width for the approximate Gaussian fit.}
\label{figure35}
\end{figure*}

\begin{figure*}[ht] \centerline{\includegraphics[width=0.800\textwidth]
{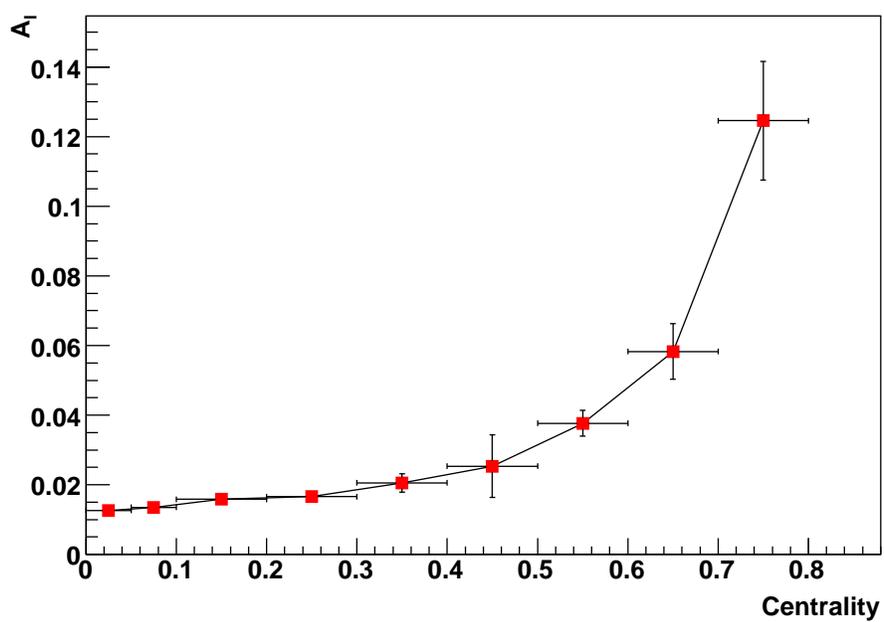}} \caption[]{``(Color online)'' The LS signal amplitude for the
large Gaussian in the fit.}
\label{figure36}
\end{figure*}
                                                                         
\begin{figure*}[ht] \centerline{\includegraphics[width=0.800\textwidth]
{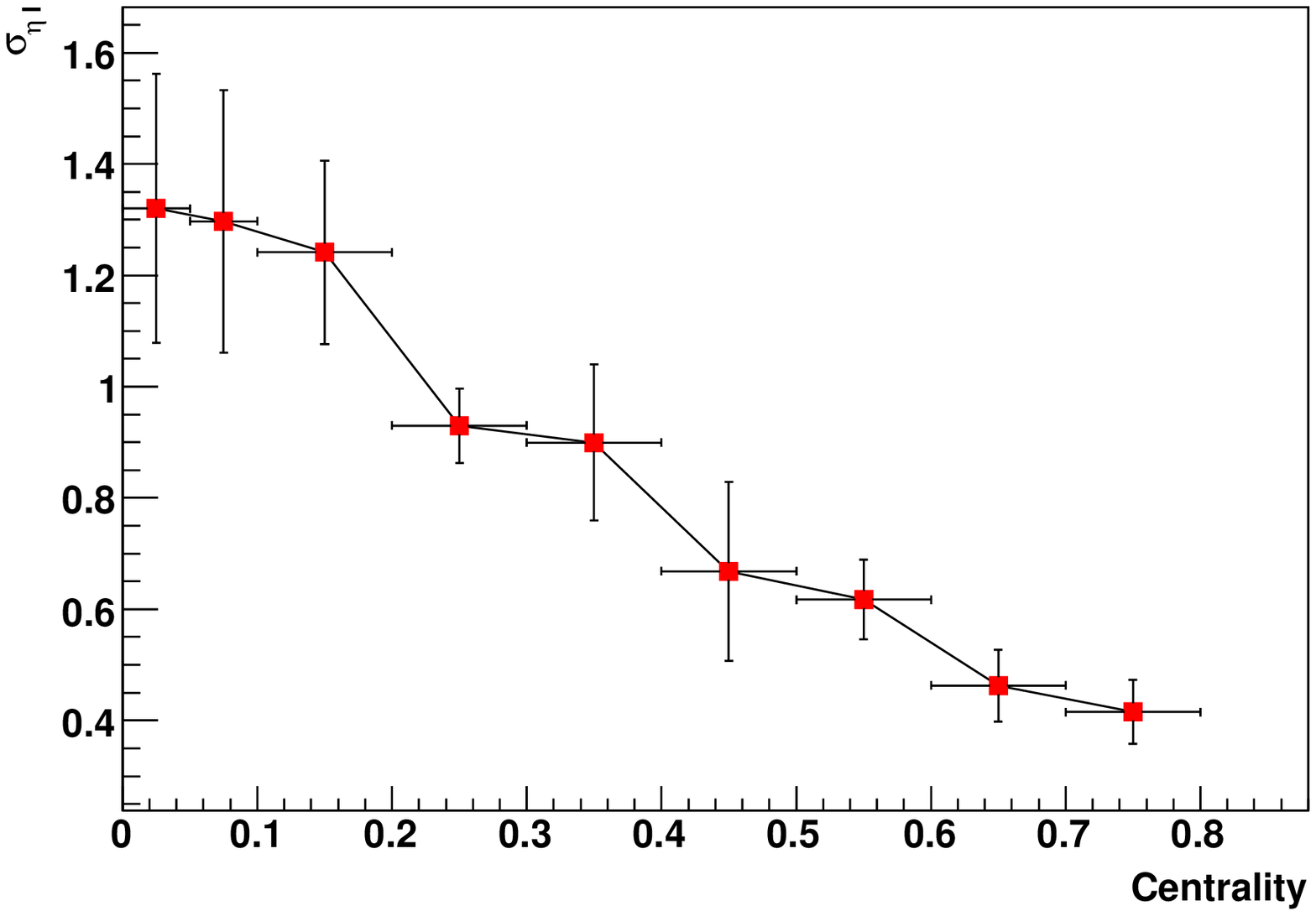}} \caption[]{``(Color online)'' The LS signal $\Delta \eta$
width for the large Gaussian in the fit.}
\label{figure37}
\end{figure*}
                                                                      
\begin{figure*}[ht] \centerline{\includegraphics[width=0.800\textwidth]
{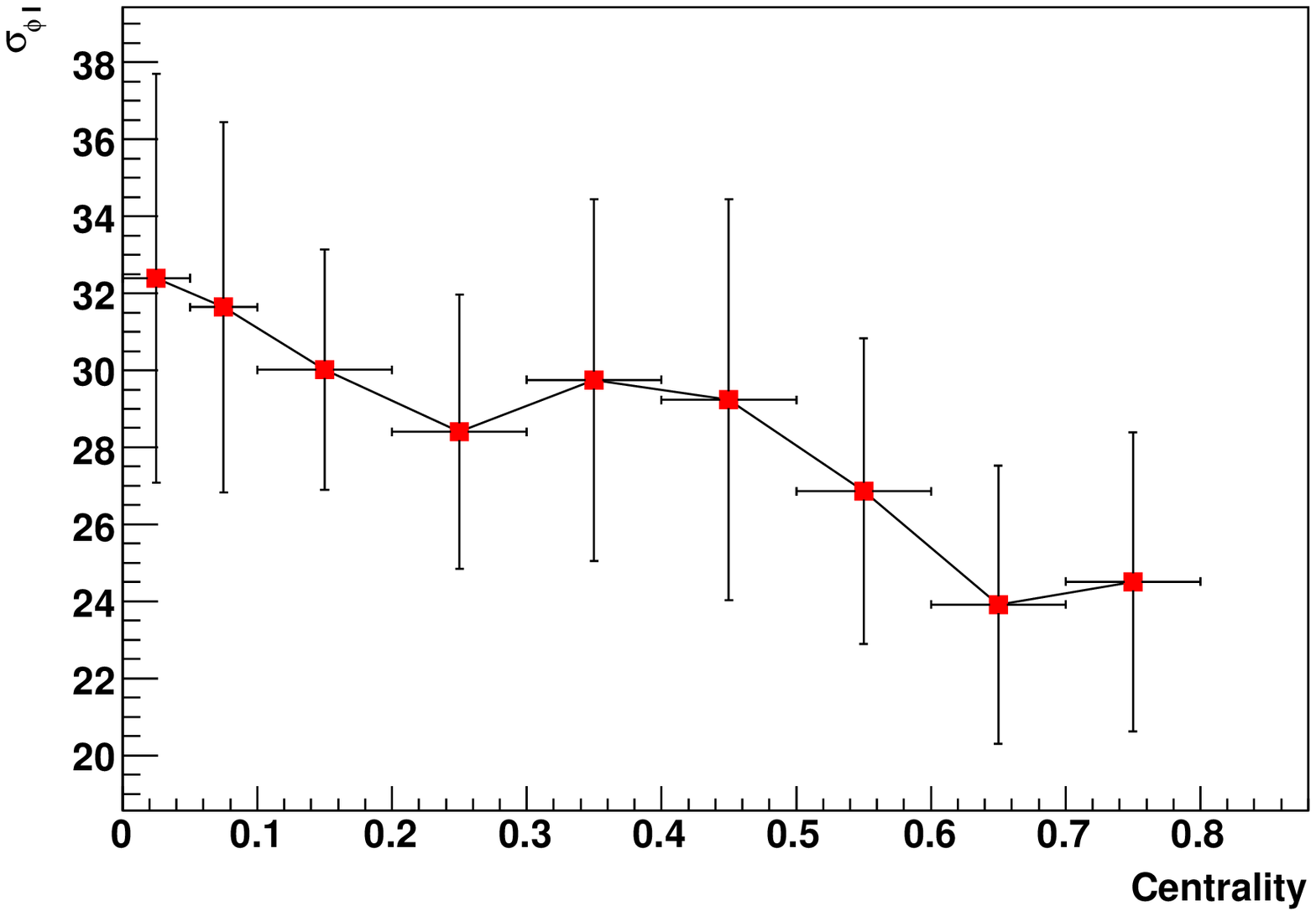}} \caption[]{``(Color online)'' The LS signal $\Delta \phi$
width for the large Gaussian in the fit.}
\label{figure38}
\end{figure*}

\clearpage

\begin{figure*}[ht] \centerline{\includegraphics[width=0.800\textwidth]
{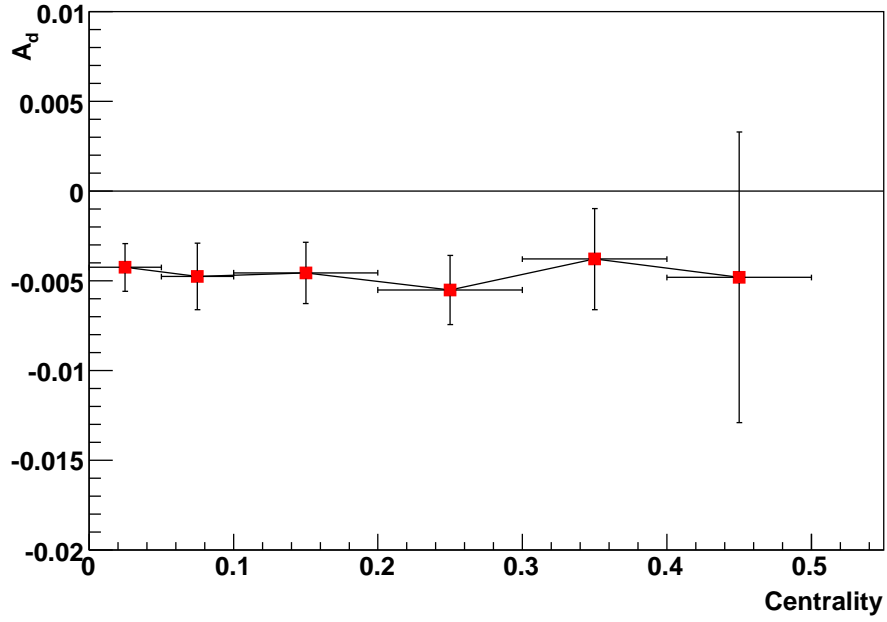}} \caption[]{``(Color online)'' The LS signal amplitude for the
dip Gaussian in the fit.}
\label{figure39}
\end{figure*}
                                                                           
\begin{figure*}[ht] \centerline{\includegraphics[width=0.800\textwidth]
{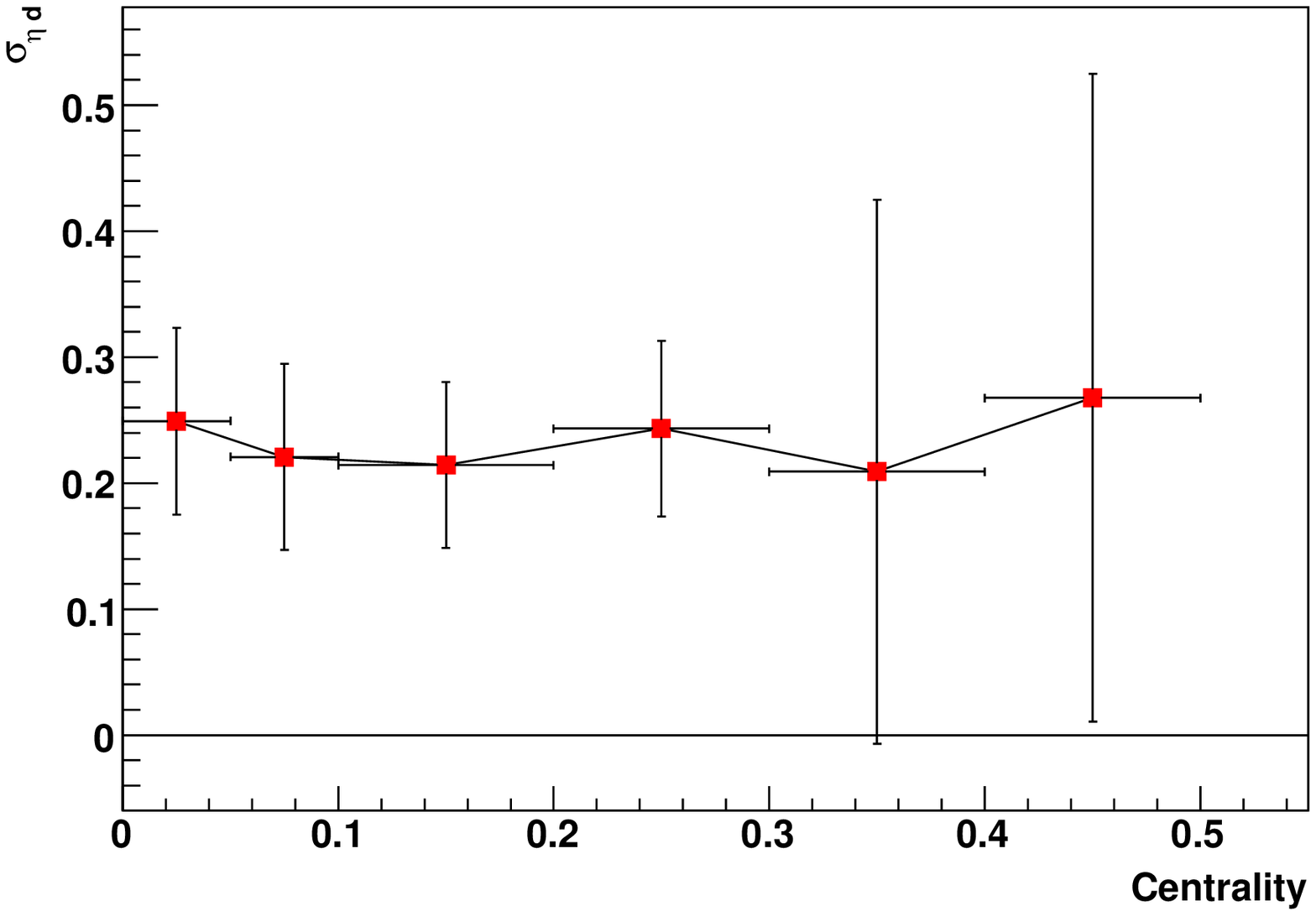}} \caption[]{``(Color online)'' The LS signal $\Delta \eta$
width for the dip Gaussian in the fit.}
\label{figure40}
\end{figure*}
                                                        
\begin{figure*}[ht] \centerline{\includegraphics[width=0.800\textwidth]
{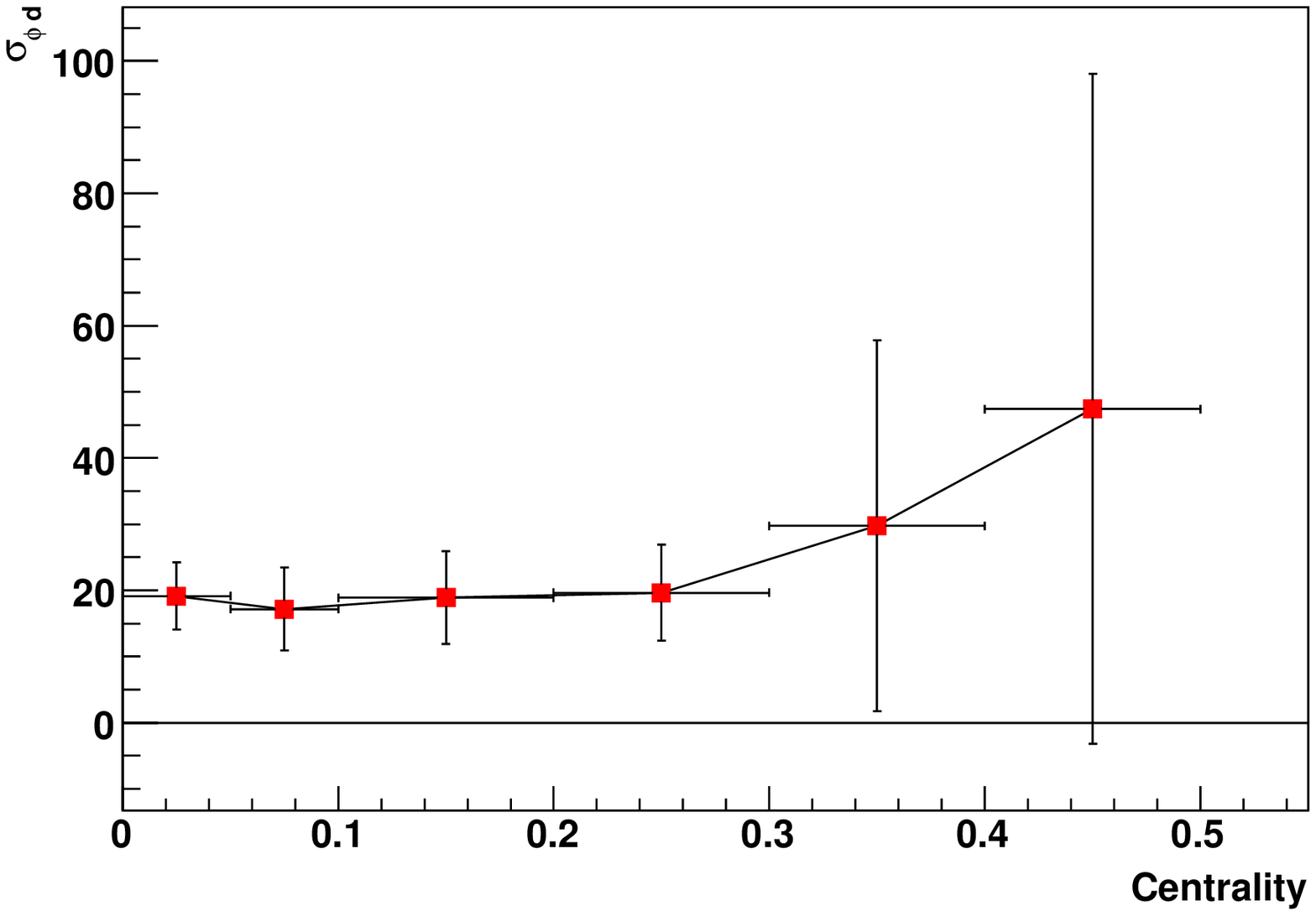}} \caption[]{``(Color online)'' The LS signal $\Delta \phi$
width for the dip Gaussian in the fit.}
\label{figure41}
\end{figure*}

\end{document}